\documentclass[prx,twocolumn,superscriptaddress,secnumarabic,balancelastpage]{revtex4-1}
\usepackage{amssymb}
\usepackage{amsmath,amsthm}
\usepackage[dvipsnames]{xcolor}
\usepackage{chngpage}
\usepackage{lgrind}        % convert program listings to a form includable in a LaTeX document
\usepackage{color}         % produces boxes or entire pages with colored backgrounds
\usepackage{graphics}      % standard graphics specifications
\usepackage[pdftex]{graphicx}      % alternative graphics specifications
\usepackage{longtable}     % helps with long table options
\usepackage{epsf}          % old package handles encapsulated post script issues
\usepackage{bm}            % special 'bold-math' package
\usepackage{asymptote}     % For typesetting of mathematical illustrations
\usepackage{thumbpdf}
\usepackage{verbatim}
\usepackage{url}
\usepackage{MnSymbol}
\usepackage{physics}
\usepackage[colorlinks=true]{hyperref} 
\usepackage{enumitem}	% compact enumeration

\usepackage{float}
\usepackage{times}
\usepackage{color}
\usepackage{verbatim}
\usepackage{soul,xcolor}
\setstcolor{red}
\usepackage{colortbl}
\usepackage[normalem]{ulem} % For \sout (and hence \cut)
\usepackage{tabularx}
\usepackage{bbold}
\newtheorem{theorem}{Theorem}[section]

\makeatletter
\AtBeginDocument{
\g@addto@macro{\appendix}{\renewcommand{\p@subsection}{\@Alph\c@section}}
}
\makeatother

\begin{document}

\title{Robust Hamiltonian Engineering for Interacting Qudit Systems}

\affiliation{Department of Physics, Harvard University, Cambridge, Massachusetts 02138, USA}
\affiliation{School of Engineering and Applied Sciences, Harvard University, Cambridge, Massachusetts 02138, USA}

\author{Hengyun Zhou$^1$}
\thanks{These authors contributed equally to this work}
\author{Haoyang Gao$^1$}
\thanks{These authors contributed equally to this work}
\author{Nathaniel T. Leitao$^1$}
\author{Oksana Makarova$^{1,2}$}
\author{Iris Cong$^1$}
\author{Alexander M. Douglas$^1$}
\author{Leigh S. Martin$^1$}
\author{Mikhail D. Lukin$^1$}
%\email{Corresponding email: lukin@physics.harvard.edu}

% Include the date command, but leave its argument blank.

\begin{abstract}
We develop a formalism for the robust dynamical decoupling and Hamiltonian engineering of strongly interacting qudit systems.
Specifically, we present a geometric formalism that significantly simplifies qudit pulse sequence design while incorporating 
the necessary robustness conditions.
We experimentally demonstrate these techniques in a strongly-interacting, disordered ensemble of spin-1 nitrogen-vacancy centers, achieving over an order of magnitude improvement in coherence time over existing pulse sequences.
We further describe how our techniques enable the engineering of exotic many-body phenomena such as quantum many-body scars, and allow enhanced sensitivities for quantum metrology.
These results enable the engineering of a whole new class of complex qudit Hamiltonians, with wide-reaching applications in dynamical decoupling, many-body physics and quantum metrology.
\end{abstract}
\maketitle

\section{Introduction}
The design and implementation of novel Hamiltonians opens up a wide range of opportunities in quantum science and technology.
Examples range from one-axis twisting Hamiltonians for entanglement-enhanced quantum metrology~\cite{wineland1992spin,kitagawa1993squeezed,ma2011quantum}, the toric code Hamiltonian for quantum computation~\cite{kitaev2003fault,satzinger2021realizing,semeghini2021probing}, to various XXZ spin chain models for quantum many-body physics~\cite{hild2014far,jepsen2020spin,wei2021quantum}.
One approach to the experimental implementation of such models is to build specific quantum simulator systems, where the desired Hamiltonian is directly realized in the system~\cite{bloch2012quantum,georgescu2014quantum,altman2021quantum}.
An alternative approach is to start with the native Hamiltonian of a system, and employ so-called Hamiltonian engineering techniques to transform this native Hamiltonian into a desired form~\cite{eckardt2017colloquium,bukov2015universal}.
Such methods broadly fall under the moniker of Floquet engineering, and have emerged as a powerful way to turn a quantum simulator of one specific Hamiltonian into a simulator of many desired systems~\cite{goldman2014periodically,choi2020robust,wei2018exploring,geier2021floquet}.
As a special case of Hamiltonian engineering, dynamical decoupling of interactions~\cite{waugh1968approach,ajoy2019selective,choi2017dynamical,burum1979analysis,cory1990time} plays a particularly important role, both in preserving the state of the system when needed, and as a key step  towards the engineering of more complex interaction Hamiltonians.

\begin{figure}
\begin{center}
\includegraphics[width=\columnwidth]{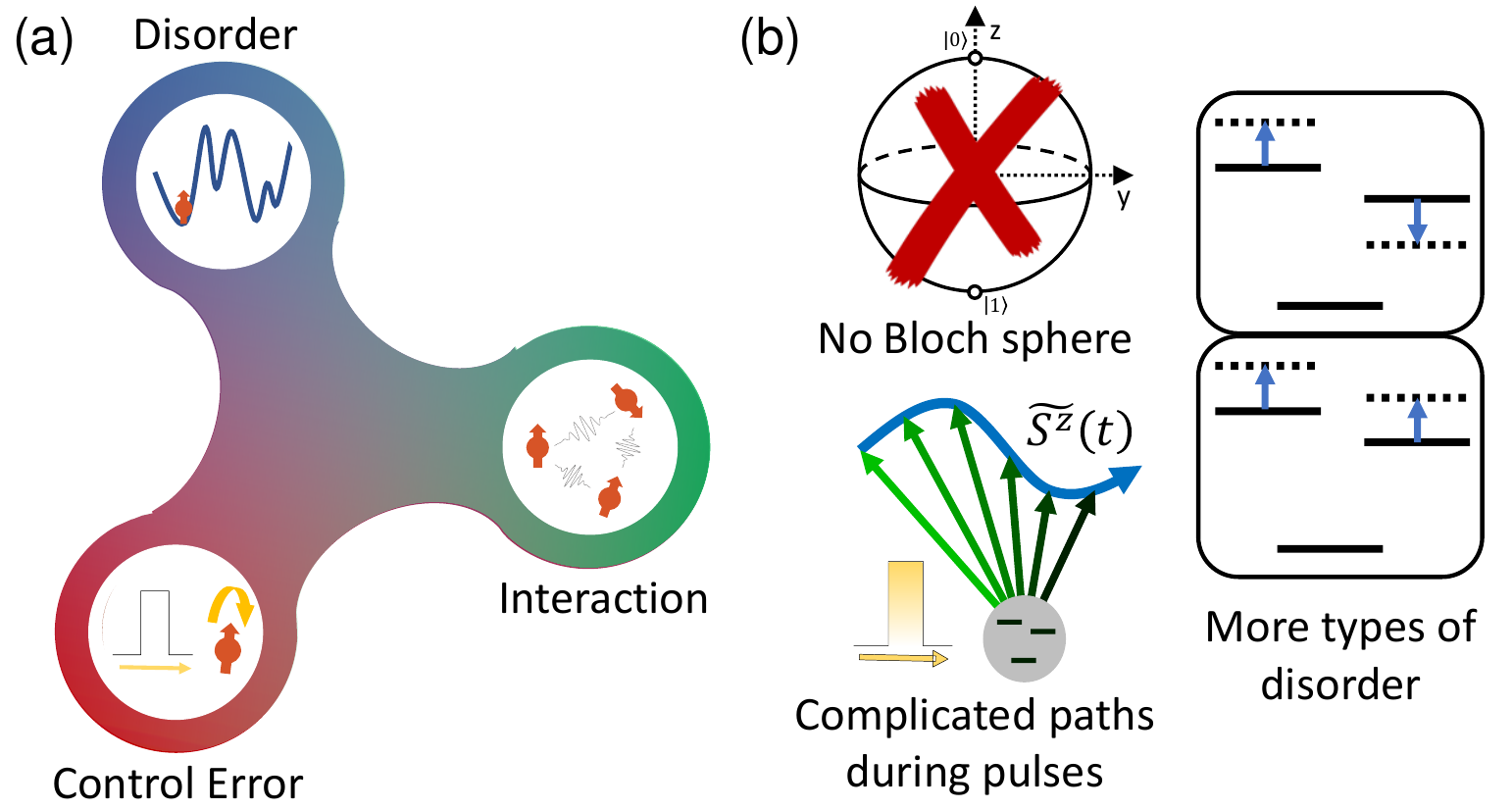}
\caption{{\bf Problem overview.} (a) Overview of the robust Hamiltonian engineering problem. Our goal is to transform the interaction and on-site disorder in a desired way, while also being insensitive to the effect of various control errors. (b) Challenges in the qudit case, including the lack of a Bloch sphere picture, more independent types of on-site disorder that require simultaneous cancellation, and a more complicated trajectory of the transformed $S^z$ operator during pulses (color gradient represents time during pulses, see Sec.~\ref{sec:robusthigherspin} and SI Sec.~S2.D for discussions).} 
\label{fig:intro}
\end{center}
\end{figure}

Until now, the majority of existing Hamiltonian engineering methods for spin systems have focused on qubits, due to their ease of manipulation, the availability of geometric intuition from the Bloch sphere, relevance to many experimental systems, as well as maturity of control techniques originally developed in the nuclear magnetic resonance (NMR) community~\cite{haeberlen1968coherent,waugh1968approach,vandersypen2001experimental,hahn1950spin,viola1999dynamical,burum1979analysis,cory1990time}.
Extending these techniques to qudit systems with more than two states, however, presents several new opportunities~\cite{choi2017dynamical}.
For quantum many-body physics, qudits enable a richer landscape of Hamiltonians, allowing for new explorations of  quantum many-body scars~\cite{schecter2019weak}, quantum chaos~\cite{blok2021quantum}, and additional spin-exchange channels~\cite{davis2019photon,stamper-kurn2013spinor}.
In quantum metrology, the larger spin results in a larger dipole moment for enhanced sensitivity~\cite{fang2013high,mamin2014multipulse,bauch2018ultralong}, and may also allow time-reversal operations that are not readily accessible with two levels~\cite{choi2017dynamical,davis2016approaching,hosten2016quantum}.
Moreover, such techniques will also be relevant for a large number of experimental platforms, including nitrogen-vacancy centers in diamond (spin-1)~\cite{fang2013high,mamin2014multipulse,bauch2018ultralong,kucsko2018critical}, quadrupolar NMR ($^2$H, $^{14}$N have nuclear spin-1)~\cite{vega1976fourier,brauniger2013solid,Chandrakumar1996}, cold molecules~\cite{bohn2017cold,lepoutre2019out}, and nuclear spins or hyperfine states in trapped atoms and ions~\cite{patscheider2020controlling,gorshkov2010two,zhang2014spectroscopic,gabardos2020relaxation,davis2020protecting}.

Designing qudit Hamiltonian engineering sequences, however, is challenging (see Fig.~\ref{fig:intro}(b)).
The significantly larger Hilbert space leads to many more types of interactions, disorder and error channels, and  control is often available only on a subset of transitions due to selection rules.
At the same time, the lack of a simple Bloch sphere picture makes the design procedure much less intuitive~\cite{kemp2021nested,barnett2006classifying,makela2007inert,serrano-ensastiga2020majorana,ribeiro2007thermodynamical,giraud2015tensor}.
Moreover, even if it were possible to design sequences to engineer Hamiltonians in the case of ideal pulses, it is not clear whether they could be made robust to experimental imperfections such as finite pulse durations and other pulse errors.
Indeed, prior proposals for qudit Hamiltonian engineering and interaction decoupling~\cite{choi2017dynamical,okeeffe2019hamiltonian} do not take any practical robustness considerations into account, making their practical applications 
challenging (see e.g. Fig.~\ref{fig:Decoupling_experiment_results}). 
Consequently, while there has been work on single qudit dynamical decoupling~\cite{vega1976fourier,brauniger2013solid,Chandrakumar1996,vitanov2015dynamical,yuan2022preserving}, up to now there have been no experimental demonstrations of full disorder and interaction decoupling for interacting spin systems with more than two levels.

In this work, we develop a general formalism for qudit Hamiltonian engineering, and use this to design and, for the first time,  experimentally demonstrate practical decoupling of spin-1 dipolar interactions. At the same time, our sequence  also decouples on-site disorder, and achieves robustness against control errors and disorder during pulses (Fig.~\ref{fig:intro}(a)).
More specifically, motivated by recent advances in the robust Hamiltonian engineering of disordered and interacting qubit systems~\cite{choi2020robust}, we devise a representation of qudit Hamiltonian transformations based on the interaction picture transformations of the $S^z$ operator, for any secular interaction Hamiltonian  satisfying the rotating wave approximation (RWA).
We find that the implementation of such transformations and corresponding analysis of finite pulse duration effects and other imperfections can be easily achieved with a graphical representation of the desired transformations, where pulse sequences represent a walk through the graph that starts and ends at the same node (Fig.~\ref{fig:Main_Results}(d-e)).
Using these insights, we focus on the challenge of designing robust disorder and interaction decoupling sequences for an ensemble of interacting spin-1 nitrogen vacancy (NV) centers in diamond, where only magnetically-allowed transitions can be driven.
We design several classes of such pulse sequences, and experimentally realize a significant improvement in coherence time over qutrit pulse sequences that only decouple disorder, representing the first demonstration of full disorder and interaction decoupling in a qudit system.

The ability to robustly engineer qudit Hamiltonians represents an important step towards the realization of complex interaction Hamiltonians for quantum many-body physics and quantum metrology, and we describe how our techniques can be employed in these applications.
As a demonstration of the rich landscape of Hamiltonians now accessible in qudit systems, we devise pulse sequences that transform the native NV-NV interaction between two groups of NVs with different lattice orientations into a spin-1 XY Hamiltonian, realizing an exotic bipartite quantum many-body scar~\cite{schecter2019weak}.
For quantum metrology, we discuss how higher spin systems naturally lead to an enhanced effective dipole moment for magnetic field sensing, and how to maximize sensitivity given the complicated transformations enacted by the Hamiltonian engineering~\cite{pang2017optimal}.

This paper is organized as follows: 
In Sec.~\ref{sec:main_results}, we summarize the main achievements and key techniques developed in this work.
In Sec.~\ref{sec:formalism}, we introduce our general formalism for designing robust sequences in qudit systems, focusing on three key insights that enable robust sequence design.
In Sec.~\ref{sec:decoupling}, we analyse a specific example of qutrit decoupling sequence design, and in Sec.~\ref{sec:experiment} we demonstrate experimentally significant improvements in decoupling performance over existing pulse sequences.
We then apply these techniques to quantum many-body physics and quantum metrology in Sec.~\ref{sec:scars} and Sec.~\ref{sec:sensing}, and conclude in Sec.~\ref{sec:conclusions} with an outlook for  future directions.

\section{Main Results}
\label{sec:main_results}
\begin{figure*}
\begin{center}
\includegraphics[width=2\columnwidth]{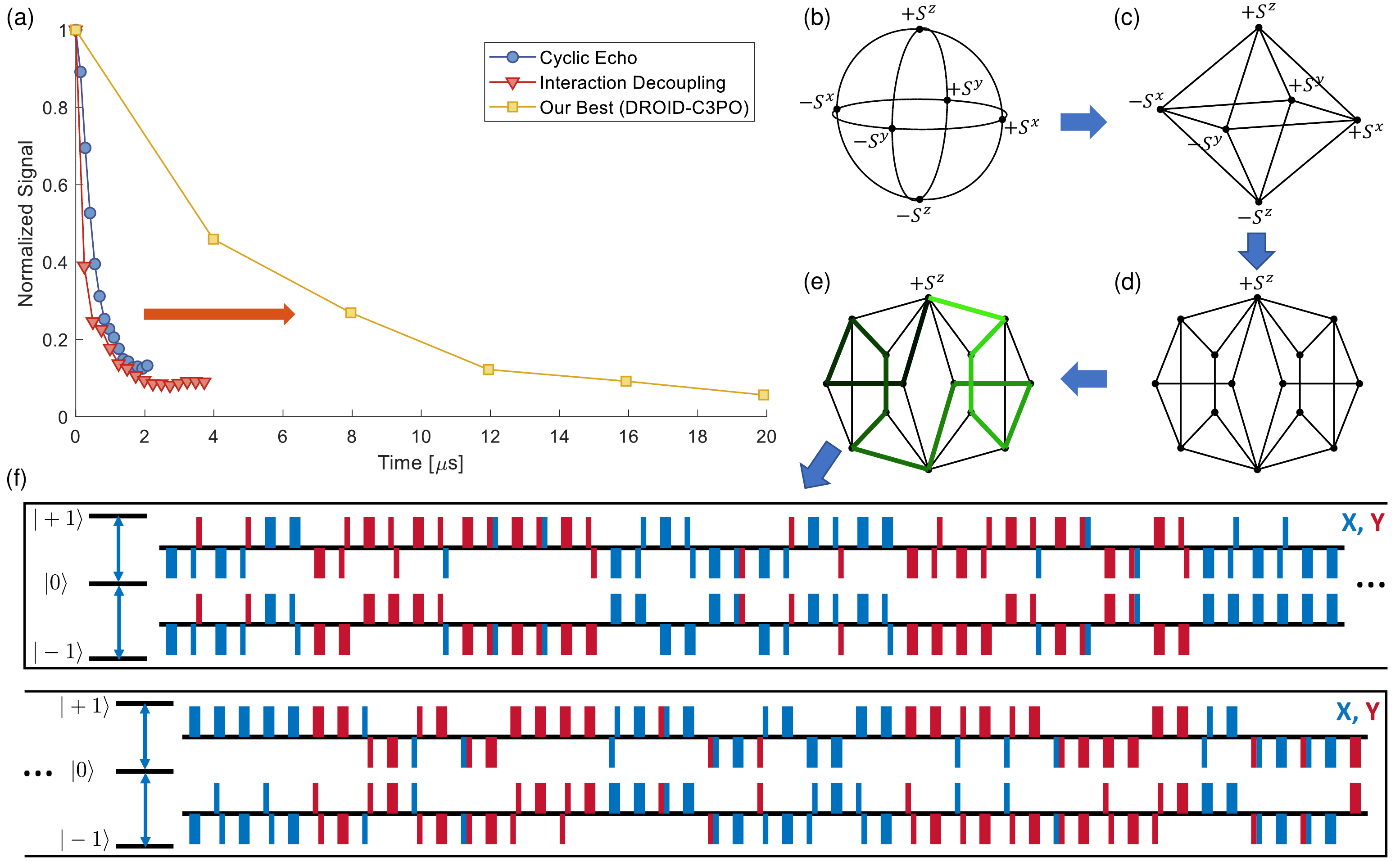}
\caption{{\bf Summary of main results.} (a) Experimental demonstration of an order of magnitude improvement in decoupling timescale compared to existing sequences. The plotted curve is the measured average decay trace for different pulse sequences, where the average is taken over all three coherent superposition initial states $\frac{\ket{0}+\ket{-1}}{\sqrt{2}}$, $\frac{\ket{0}+\ket{+1}}{\sqrt{2}}$, and $\frac{\ket{+1}+\ket{-1}}{\sqrt{2}}$.
(b-c) After trivially replacing all arcs in a qubit Bloch sphere by straight lines, we get an octahedron, which is a special example of our new concept ``decoupling frame graphs". (d) A generic decoupling frame graph, where each vertex represents a distinct frame and each edge represents a pair of frames connected by an experimentally implementable pulse. (e) A pulse sequence is then represented by a walk traversing all the vertices. The color gradient indicates the order of the walk, and the walk is translated into a pulse sequence that can be implemented in experiments. (f) Plot of our current best qutrit decoupling sequence ``DROID-C3PO" (i.e. Disorder-RObust Interaction Decoupling - Coherent 3-level Pulse Optimization). All pulses in this pulse sequence simultaneously drive the two transitions with equal amplitude. The thin lines represent spin-1 $\pi/2$ pulses (i.e. rotation of the spin-1 generalized Bloch sphere by an angle $\pi/2$, experimentally implemented by simultaneously driving the two transitions with two $\frac{\pi}{\sqrt{2}}$ pulses) and the thick lines represent spin-1 $\pi$ pulses. The color of the pulses represent the pulse axis (X or Y), and the direction of the pulses (up or down) represent the two opposite rotation directions (e.g. $+\pi/2$ pulse and $-\pi/2$ pulse). The proportions of this plot are drawn consistently with actual time durations. The ellipsis in the plot indicates that the two rows are connected.}
\label{fig:Main_Results}
\end{center}
\end{figure*}

The most important result in this work is the design and realization of a robust disorder and interaction decoupling pulse sequence for an interacting qudit ensemble. By building in robustness against various control imperfections, \textbf{we experimentally demonstrate an order of magnitude improvement in the decoupling timescale compared to existing sequences}~\cite{choi2017observation,choi2017dynamical}. This improvement is shown in Fig.~\ref{fig:Main_Results}(a) and the pulse sequence we designed is plotted in Fig.~\ref{fig:Main_Results}(f).

In order to design these qudit pulse sequences, we examine the transformed Hamiltonian in the interaction picture $\tilde{H}=U^\dagger HU$ with respect to the control unitaries $U(t)$. We will now describe two key ideas behind our systematic design approach.
First, to overcome the lack of a Bloch sphere intuition in the qudit case, we propose a new graphical representation that generalizes the Bloch sphere representation for qubit pulse sequences.
Specifically, in this approach:
\begin{itemize}
    \item The Bloch sphere is generalized into a new concept that we call the ``decoupling frame graph" (Fig.~\ref{fig:Main_Results}(b-e)), in which each vertex represents a distinct frame (defined as the transformed higher spin $S^z$ operator $\tilde{S}^z=U^\dagger S^zU$), leading to a distinct transformed Hamiltonian, and each edge represents an experimentally implementable pulse connecting the two frames on its ends.
    \item A pulse sequence is equivalent to a path on the graph that traverses a set of vertices and edges.
    \item This representation only keeps track of $\tilde{S}^z$, which we show to be sufficient to describe the transformed Hamiltonian as long as the native Hamiltonian is secular (i.e. satisfies the RWA approximation, see Theorem~\ref{thm:secular}).
\end{itemize}

Second, for practical experimental implementations, it is crucial that the pulse sequence is robust against undesired dynamics during the pulses~\cite{choi2020robust,zhou2020quantum}. We achieve this by identifying particularly simple trajectories for the transformed $\tilde{S}^z$, such that we can easily analyze the frames during pulses as well. Specifically, 
\begin{itemize}
    \item We focus on pulses that transform $S^z$ along geodesics on the generalized Bloch sphere~\cite{choi2017dynamical,macfarlane1968description}, since these trajectories can be decomposed as a linear combination of the frames before and after the pulse (see Fig.~\ref{fig:geodesic_consideration}(b)). This property simplifies the robustness condition from a property of the whole trajectory into a property of a few discrete frames. Moreover, it allows us to elegantly cancel certain terms by simply going through pairs of antipodal points on the generalized Bloch sphere, as we discuss in more detail in Sec.~\ref{sec:decoupling}. This is also the key property that enabled the design of robust qubit pulse sequences in previous work~\cite{choi2020robust}.
    \item In the qutrit case, geodesic trajectories can be guaranteed by using balanced double driving pulses (i.e. pulses that drive the $\ket{0}\leftrightarrow\ket{+1}$ and $\ket{0}\leftrightarrow\ket{-1}$ transitions simultaneously with equal amplitude), generalizing the ``great arc" trajectory on a spin-1/2 Bloch sphere during $\pi/2$ pulses.
    \item We exploit additional similarities between the qutrit and qubit cases, including related structures of the decoupling frame graphs (see Fig.~\ref{fig:Two_blocks}), and similarities between cancelling $\left(S^z\right)^2$ disorder in the spin-1 case and cancelling the spin-$\frac{1}{2}$ Ising interaction.
    \end{itemize}
    
Combining these insights and exploiting certain structures from higher-order qubit sequences~\cite{zhou2023robust}, we construct a qutrit decoupling sequence which is not only robust, but also naturally inherits further higher-order performance improvements. This enables the significant extensions in coherence time experimentally demonstrated in Sec.~\ref{sec:experiment} and Fig.~\ref{fig:Main_Results}(a). Finally, we also show that our formalism can be utilized to engineer a quantum many-body scar Hamiltonian (Sec.~\ref{sec:scars}), as well as to design sensing sequences for higher spin sensors that promise enhanced sensitivity (Sec.~\ref{sec:sensing}).

\section{General Formalism for Qudit Hamiltonian Engineering}
\label{sec:formalism}
In this section, we will introduce our general formalism for robust qudit Hamiltonian engineering.
Key to our formalism are insights into compact algebraic and graphical representations of the engineered Hamiltonian, combined with judicious choices of pulse families to satisfy real-world constraints and achieve robustness.
Many of these observations are inspired by methods for robust qubit sequence design, yet require viewing the results from a new perspective and making substantial generalizations.
For a more detailed description of robust qubit sequence design, we refer the reader to Ref.~\cite{choi2020robust}.

In each subsection below, we will first illustrate the intuitions behind key insights using simple examples with qubits, and then generalize the statements to the qudit case.
As we shall see, this is a nontrivial extension, and requires developing new geometric intuitions and understandings of the Hamiltonian engineering constraints.

\subsection{Hamiltonian Representation and Decoupling Frame Set}
\label{sec:higherspinframes}
The setting we are interested in is a generic qudit Hamiltonian
\begin{equation}
    H=\sum_i H_i^{\textrm{dis}} + \sum_{i\neq j} H_{ij}^{\textrm{int}},
\end{equation}
where the first term describes an on-site disorder, and the second term describes a symmetric two-body interaction that satisfies the rotating wave approximation (i.e. the secular approximation).
We focus on the case where we only have global control over the spin system, consisting of pulses allowed by the selection rules of the system.
Going into the interaction picture with respect to the ideal pulses, we can write the interaction picture Hamiltonian as $\tilde{H}_k=U_{k-1}^\dagger HU_{k-1}$ with $U_{k-1}=P_{k-1}\cdots P_1$. In average Hamiltonian theory, the evolution of the system can be described by an effective Hamiltonian $H_{\textrm{eff}}$, which is the average of $\tilde{H}_k$ weighted by the corresponding evolution times $\tau_k$~\cite{haeberlen1968coherent}, $H_{\textrm{eff}}=\sum_k \tau_k\tilde{H}_k/T$, $T$ being the Floquet period.
The goal of Hamiltonian engineering is then to design a pulse sequence $\{P_k\}$ that leads to the desired form of $H_{\textrm{eff}}$.

The key insight that significantly simplifies the pulse sequence design problem is that the Hamiltonian transformation $\tilde{H}=U^\dagger HU$ is uniquely determined by the transformations of the higher spin $S^z$ operator, $\tilde{S}^z=U^\dagger S^z U$. 
This observation allows us to keep track of only the $S^z$ transformation, instead of the whole unitary $U$ or the pulse sequence history, which contain unnecessary information about the transformations of $S^x$ and $S^y$.
As an example of this observation, let us examine the example of a dipolar-interacting spin-$\frac{1}{2}$ system:
\begin{equation}
    H=\sum_i h_iS_i^z+\sum_{ij}J_{ij}\left(S_i^x S_j^x+S_i^y S_j^y-2S_i^z S_j^z\right).
\label{eq:Hspin1/2}
\end{equation}
Because the on-site disorder term is proportional to $S^z$, it is obvious that its transformation is determined by $\tilde{S}^z$.
To see that this is also the case for the interaction term, notice that the interaction term can be rewritten as $\vec{S}_i\cdot\vec{S}_j-3S_i^zS_j^z$, where the first term is spherically-symmetric and therefore invariant under global qubit rotations $\hat{U}$, and the second term is manifestly determined by $\tilde{S}^z$.

The fundamental reason for this insight is the rotating wave approximation (i.e. secular approximation).
Intuitively, when there is a strong quantizing field that separates the two energy levels, the Hamiltonian rapidly rotates around the $z$ axis.
Therefore, any part of the Hamiltonian that is not rotationally invariant rapidly averages out, and the resulting Hamiltonian (e.g. Eq.~(\ref{eq:Hspin1/2})) must be invariant under $z$ rotations.
Because of this, the rotation of $x$ and $y$ axes in the plane perpendicular to the $z$ axis does not matter, as they are equivalent, and the transformation of $z$ axis determines everything.

The observation that $\tilde{S}^z$ uniquely determines $\tilde{H}$ can be directly generalized into the higher spin case, and we formulate this statement precisely as the following theorem:
\begin{theorem}
For two unitaries $U_1$, $U_2$, such that $\tilde{S}^z=U_1^\dagger S^z U_1=U_2^\dagger S^z U_2$, we have $(U_1^\dagger)^{\otimes n} H U_1^{\otimes n}=(U_2^\dagger)^{\otimes n} H U_2^{\otimes n}$, where $n$ is the number of spins in the system and $H$ is a qudit Hamiltonian satisfying the secular approximation on each transition.
\label{thm:secular}
\end{theorem}
The full proof is given in SI Sec.~S1.C, but the basic intuition is similar to the spin-$\frac{1}{2}$ case: the rotating wave approximation causes operators other than diagonal ones to drop out, leaving $\tilde{S}^z$ as the only relevant information.

Since transformations of the $S^z$ operator alone are sufficient to describe the Hamiltonian transformations, we only need to keep track of them for the Hamiltonian engineering problem.
From now on, we will refer to the transformation of the $S^z$ operator as a ``frame" or ``frame transformation". Compared to prior approaches that did not utilize the secularity of the Hamiltonian~\cite{choi2017dynamical,okeeffe2019hamiltonian}, Theorem~\ref{thm:secular} reduces the information we need to keep track of, and makes it possible to extract physical insights from the frames $\tilde{S}^z$ themselves that help inform sequence design, as we shall see in Sec.~\ref{sec:decoupling}.

The proof in SI Sec.~S1.C does not provide an explicit construction of the transformed Hamiltonian as a function of $\tilde{S}^z$, although it is easy to show that the final result will be a polynomial in the decomposition coefficients of $\tilde{S}^z$ in the Gell-Mann basis, as we describe in SI Sec.~S1.D.
Using the representation theory of Lie groups, such expressions can in fact also be directly constructed, as we show in complementary work~\cite{leitao2022qudit}.

\begin{figure*}
\begin{center}
\includegraphics[width=2\columnwidth]{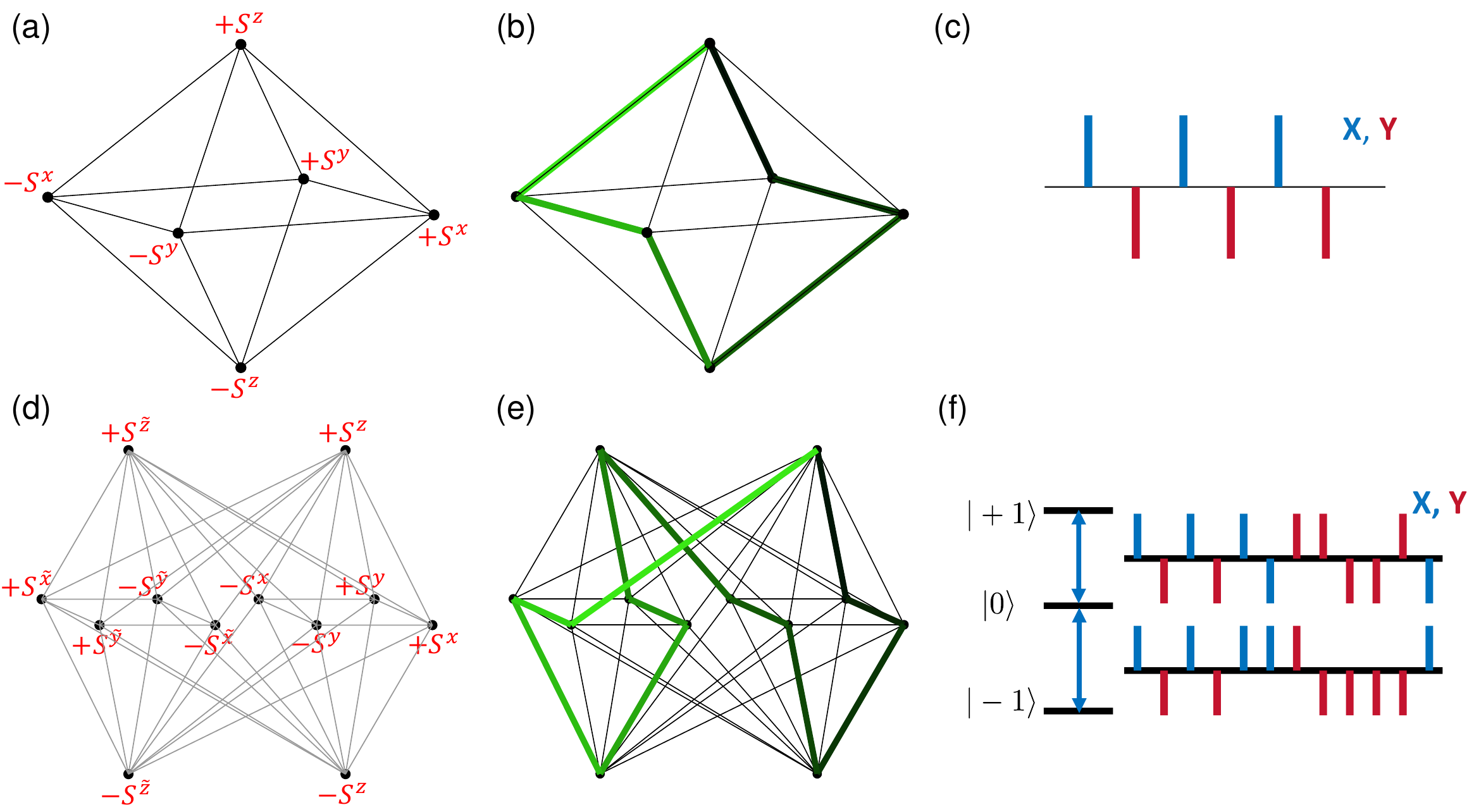}
\caption{{\bf Decoupling frame graphs.}  Geometric representation of decoupling pulse sequences. Vertices indicate transformed $S^z$ frames and edges indicate physically-implementable pulses that connect different frames.
(a-c) Representative qubit decoupling sequence. 
Subplot (a) shows the decoupling frame graph for the qubit case, and a walk on it (b, walk ordering given by color gradient) produces a pulse sequence consisting of $\pi/2$ pulses (c).
(d-f) Corresponding results for a qutrit decoupling sequence. These plots are only for high level illustration of our methods; the details of qutrit decoupling sequence design are discussed in Sec.~\ref{sec:decoupling}, and the precise definition of the frame set in (d) is given in Eq.~(\ref{eq:spin1_frame_set}).}
\label{fig:decoupling_frame_graph.png}
\end{center}
\end{figure*}

\subsection{Graphical Representation of Qudit Decoupling}
\label{sec:graphical}
In this section, we will introduce a new graphical representation, which we call the ``decoupling frame graph", to describe the frame set we use to decouple the interaction and the pulses connecting them.
In these graphs (e.g. Fig.~\ref{fig:decoupling_frame_graph.png}(a,d)), the vertices represent the frames we use for decoupling and the edges represent the pulses connecting these frames.
As we will discuss in this section, this graphical representation significantly simplifies the inclusion of connectivity requirements, while also providing a simple visualization of the pulse sequence.

To make Hamiltonian engineering techniques relevant to real-world experiments, it is key to ensure that the desired transformations can indeed be experimentally implemented given the constraints of selection rules. To build in robustness into a pulse sequence, the pulses connecting neighboring frames must further be simple enough that we can easily analyze the effects of disorder and interaction during them. Therefore, we require neighboring frames to be connected by simple and experimentally implementable pulses.

For qubit systems, this requirement is automatically achieved through the Bloch sphere picture Fig.~\ref{fig:Main_Results}(b), where the frames $\pm S^x, \pm S^y, \pm S^z$ are connected by simple $\frac{\pi}{2}$ pulses.
For qudits, however, existing linear programming techniques described in Ref.~\cite{choi2017dynamical,okeeffe2019hamiltonian} only consider the Hamiltonian transformations $\tilde{H}_k$ at each frame, but ignore the ordering of the frames and the pulses connecting them.
This often results in complicated composite pulses in the derived pulse sequence, which require cumbersome and structure-less algebraic simplifications and have no clear way to build in robustness.
Addressing this challenge for qudits thus requires developing new geometric approaches and intuitions, as we now describe.

Our approach is motivated by the observation that the simplicity in the qubit case comes in large part from our choice of frame sets.
Indeed, one important reason to choose the frames $\pm S^x, \pm S^y, \pm S^z$, as opposed to e.g. a tetrahedral or icosahedral basis~\cite{ben2020hamiltonian}, is that they are connected by simple $\frac{\pi}{2}$ pulses.
The direct generalization of this to the qudit case thus starts with a simple pulse set motivated by selection rules and easy analysis of robustness (Sec.~\ref{sec:robusthigherspin}), and searches for subsets of frames (which we will call the ``decoupling frame set") that achieve decoupling among the frames accessible using pulses in this pulse set.
Crucially, one should also keep track of how the different frames are connected by accessible pulses while building up the decoupling frame set.
In this way, the connectivity between the frames by simple pulses is guaranteed beforehand.

Motivated by the fact that the Bloch sphere picture in Fig.~\ref{fig:Main_Results}(b) can be viewed as a connectivity graph if the arcs are replaced by straight lines (see Fig.~\ref{fig:Main_Results}(c)), we illustrate the frame set and the connectivity between frames by a new graphical representation, which we refer to as the ``decoupling frame graph", as shown in Fig.~\ref{fig:Main_Results}(d). 
In these graphs, each vertex represents a different transformed $\tilde{S}^z$ frame, and each edge between two vertices represents a single pulse in our chosen pulse set that connects the two frames.
Using this representation, the decoupling requirement $H_{\textrm{eff}}=\sum_k \tilde{H_k}\tau_k=0$ becomes a requirement on the frame set and the time $\tau_k$ spent at each vertex, and a pulse sequence that achieves such decoupling can be represented as a path on this graph (see Fig.~\ref{fig:Main_Results}(e)), which walks along the edges to visit the desired vertices, and spends the requisite amount of time on each vertex.

A few comments are in order: first, a natural question one might have is whether the feasibility of a given edge to be experimentally implemented depends on the history of pulses applied, which could change the orientation of other operators such as $S^x$, while leaving $S^z$ invariant;
we prove in SI Sec.~S1.E that the only effect of this is to change the phase of the pulse to be applied, without changing which transitions are involved, thus not affecting the implementability of the pulses.
Second, it is usually more favorable to design graphs where all vertices used are connected in a single patch;
this eliminates the need for intermediate nodes to connect frames, which could complicate the cancellation of finite-pulse-duration effects.
Similarly, well connected graphs are preferred because they support more ways to traverse the vertices.
This extra degree of freedom can be utilized to satisfy robustness conditions.
Third, the choice of a well-motivated, implementable set of pulses is very important, as it determines the frames we consider and the connectivity between them.
As we will see in Sec.~\ref{sec:robusthigherspin}, balanced double driving pulses (i.e. pulses that simultaneously drive both $\ket{0}\leftrightarrow\ket{+1}$ and $\ket{0}\leftrightarrow\ket{-1}$ transitions with equal amplitude) are usually good choices in spin-1 systems, due to their simple transformations of disorder during pulses and their ease of implementation and calibration.

Let us now provide a few concrete examples of the decoupling frame graph to gain a bit more intuition.

First, consider the qubit case, where we would like to spend equal time along each of the 6 cardinal directions.
The vertices thus correspond to $\pm S^x$, $\pm S^y$, $\pm S^z$ frames, and the connecting edges, corresponding to $\pi/2$ pulses, organize the decoupling frame graph into an octahedron, as shown in Fig.~\ref{fig:decoupling_frame_graph.png}(a).
A representative path on this graph, as illustrated in Fig.~\ref{fig:decoupling_frame_graph.png}(b), can be directly translated into the decoupling pulse sequence shown in Fig.~\ref{fig:decoupling_frame_graph.png}(c). This sequence is a variant of the spin-1/2 WAHUHA sequence that decouples interactions and disorder~\cite{waugh1968approach}.

Another decoupling frame graph, which we use for qutrit disorder and interaction decoupling, is shown in Fig.~\ref{fig:decoupling_frame_graph.png}(d). The definition of the frames and why it decouples disorder and interactions is discussed in Sec.~\ref{sec:decoupling}; but for now, it is just an illustration of a generic qudit decoupling frame graph.
Similar to the qubit case, we can easily draw a path through all vertices in a simply-connected fashion, as illustrated in Fig.~\ref{fig:decoupling_frame_graph.png}(e).
The pulse sequence corresponding to the path is shown in Fig.~\ref{fig:decoupling_frame_graph.png}(f), and consists of balanced double driving pulses with different phases on each transition.

\subsection{Robust Qudit Decoupling}
\label{sec:robusthigherspin}
In order to incorporate robustness into sequence design, we have to analyze the transformation of the Hamiltonian during pulses.
For qudit systems, the transformation trajectory can be much more complex than the qubit case (see below and SI Sec.~S2.D), complicating robustness analysis.
Nevertheless, we will show that by carefully choosing the pulses that constitute the sequence, we can recover the favorable properties of the qubit case.

\begin{figure}
\begin{center}
\includegraphics[width=\columnwidth]{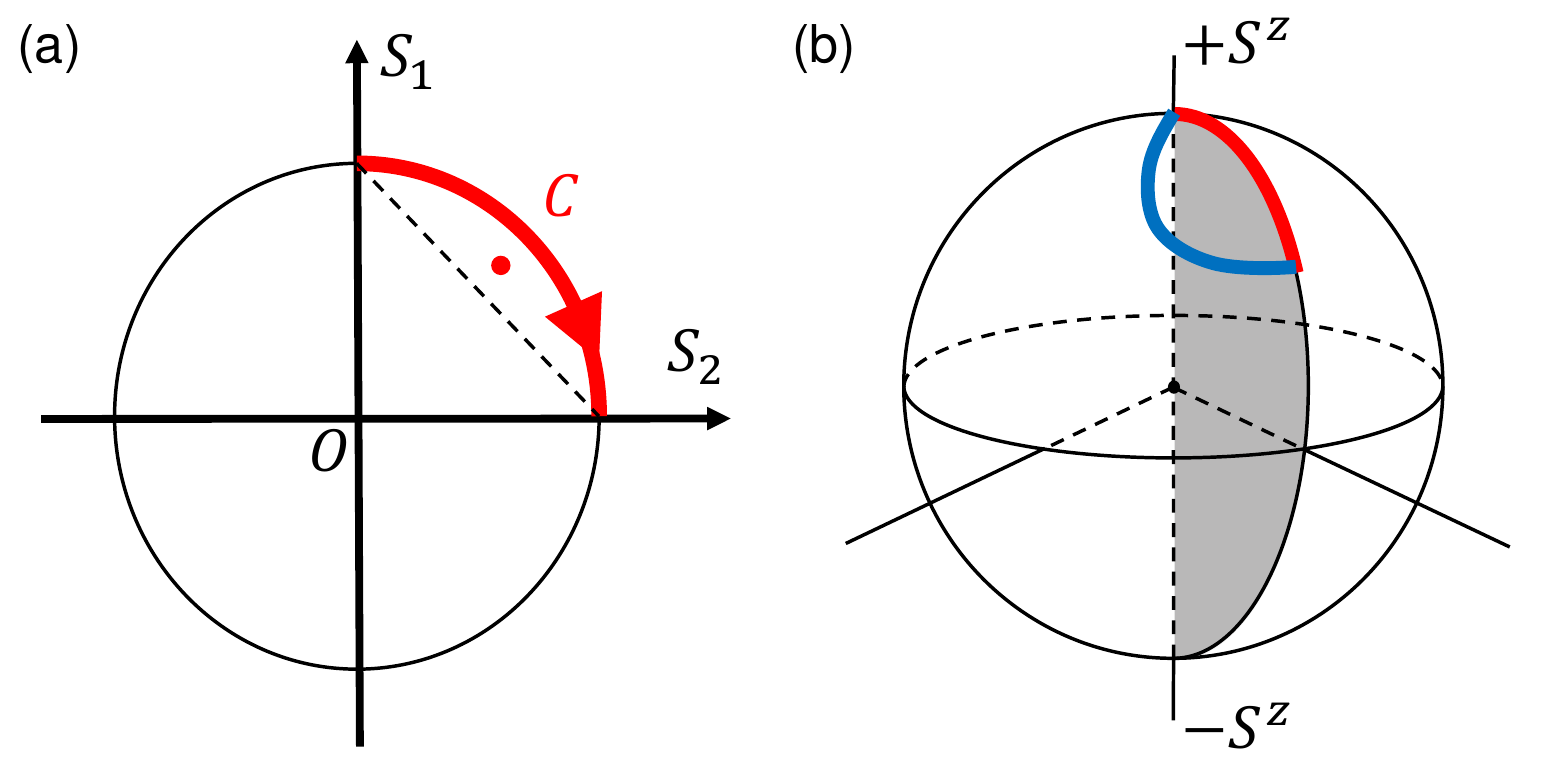}
\caption{{\bf Disorder during pulses.} (a) In the qubit case, the disorder during pulses is a continuous rotation from the frame before the pulse $S_1$ to the frame after the pulse $S_2$, as represented by the red arc. Therefore, its average effect, which is represented by the center of mass ``C" of the red arc, can be written as a scaled average of $S_1$ and $S_2$. The factor $\frac{4}{\pi}$ in Eq.~(\ref{eq:disorder_decomposition}) comes from the fact that the center of mass ``C" is slightly further from the origin compared to the midpoint of $S_1$ and $S_2$. (b) Illustration of geodesics. If the trajectory of $S^z$ during the pulse follows a geodesic, as in the case of the red curve, then the whole curve lives in the 2 dimensional subspace (i.e. the shaded plane) spanned by the frame before and after the pulse, and therefore the average effect of disorder during the pulse can naturally be decomposed as an average before and after the pulse. If the trajectory of $S^z$ does not follow a geodesic, as in the case of the blue curve, then the center of mass of the curve no longer lives in the shaded plane and no such decomposition is possible.}
\label{fig:geodesic_consideration}
\end{center}
\end{figure}

Before diving into the more complicated case of qudits, let us first briefly review how disorder during pulses is cancelled in the qubit case, to remind readers about the key properties that simplify the analysis.
Because the on-site disorder is proportional to $S^z$, we need to analyze the transformation of the $S^z$ operator during pulses.
In the qubit case, the transformation of the $S^z$ operator during pulses is a continuous rotation along a geodesic on the Bloch sphere from the frame before the pulse $S_1$ to the frame after the pulse $S_2$. If we focus on the 2 dimensional subspace that contains the trajectory of $\tilde{S}^z$ during the pulse, this trajectory is represented by the red arc in Fig.~\ref{fig:geodesic_consideration}(a). Therefore, the average effect of disorder during the pulse, which is represented by the center of mass of the red arc, can be decomposed as a simple average of $S_1$ and $S_2$. Indeed, by integrating over the pulse explicitly, one finds that the average effect of disorder during a $\pi/2$ pulse is:
\begin{equation}
    \bar{S}=\frac{4}{\pi}\left[\frac{S_1+S_2}{2}\right].
\label{eq:disorder_decomposition}
\end{equation}
The contributions of these terms can be easily incorporated into the effective Hamiltonian by treating it as extra time spent in the frames before and after the pulse, with minimal changes to the decoupling conditions otherwise.
Thus, in the qubit case, this decomposition significantly simplifies the incorporation of robustness into the sequence design problem.
For more details on qubit robust sequence design and a similar analysis of other contributions, we refer readers to Ref.~\cite{choi2020robust}.

However, this decomposition of disorder during pulses as a simple average before and after the pulse no longer holds for generic pulses in qudit case.
A concrete counterexample is the transformation of the spin-1 $S^z$ operator during a $\frac{\pi}{2}$ pulse applied on a single transition, which is discussed in more detail in SI Sec.~S2.D.
A geometric picture that explains why this decomposition (i.e. Eq.~(\ref{eq:disorder_decomposition})) no longer holds is that the trajectory of the $S^z$ transformation during pulses is no longer a geodesic on the generalized Bloch sphere for generic higher spin pulses.
To see this, examine Fig.~\ref{fig:geodesic_consideration}(b), where the red curve represents a trajectory of $\tilde{S}^z$ that follows a geodesic, and the blue curve is a trajectory that does not follow a geodesic.
For the red curve, since it lives entirely in the two dimensional subspace spanned by $\tilde{S}^z$ before and after the pulse (i.e. the shaded plane), the averaged $\tilde{S}^z$ during the pulse can always be decomposed as a scaled average of $\tilde{S}^z$ before and after the pulse.
However, for the blue curve, since its center of mass does not live in the shaded plane, there is no way to perform this decomposition.

One way to overcome this challenge is to find pulses that transform $S^z$ along geodesics, and use them to build the decoupling sequence.
For the case of a spin-1 system, we found that balanced double driving pulses (i.e. pulses that simultaneous drive both $\ket{0}\leftrightarrow\ket{+1}$ and $\ket{0}\leftrightarrow\ket{-1}$ transitions with equal amplitude) satisfy this condition.
For concreteness, let us write down the form of the Hamiltonian for balanced double driving pulses:
\begin{equation}
    H_p\propto\begin{pmatrix}
           0 & e^{-i\theta_1} & 0\\
           e^{i\theta_1} & 0 & e^{-i\theta_2}\\
           0 & e^{i\theta_2} & 0\\
           \end{pmatrix}.
\label{eq:balanced_double_driving}
\end{equation}
To see that balanced double driving pulses transform $S^z$ along geodesics, notice that the Hamiltonian of balanced double driving pulses can be related to the spin-1 $S^x$ operator by a simple conjugation $H_p=U^\dagger S^x U$, where the unitary
\begin{equation}
    U=\begin{pmatrix}
           e^{i\theta_1} & 0 & 0\\
           0 & 1 & 0\\
           0 & 0 & e^{-i\theta_2}\\
           \end{pmatrix}
\end{equation}
is a phase operator that conjugates the $S^z$ operator trivially (i.e. $U^\dagger S^z U=S^z$).
Then, the transformation of the $S^z$ operator during the pulse is:
\begin{align}
    \tilde{S}^z(t)&=e^{iH_pt}S^z e^{-iH_pt}\nonumber\\
    &= U^\dagger\left[e^{iS^xt}S^z e^{-iS^xt}\right]U.
\end{align}
Notice that the term $e^{-iS^x t}$ is a spin-1 spatial rotation operator, so its conjugation on $S^z$ transforms $S^z$ along the geodesic $\cos t~S^z+ \sin t~S^y$.
This property still holds after conjugation by $U$, and we find that for arbitrary balanced double driving pulses that rotate the spin by $\frac{\pi}{2}$, the transformation of $S^z$ during the pulse is a geodesic:
\begin{equation}
    \tilde{S}^z(\theta) = \cos\theta S_1 + \sin\theta S_2,
\label{eq:Sz_during_balanced_double_driving}
\end{equation}
where $S_{1,2}$ are the frames before and after the pulse, and $\theta$ is the angle rotated from $S_1$. We remark that the above constructions and proof can be generalized to qudits with arbitrary $d$, where the balanced double driving is generalized to a phase conjugated higher spin $S^x$ operator. Furthermore, such pulses are implementable in most experimental systems because they only require driving between neighboring $\ket{m_S}$ states.

Equation~(\ref{eq:Sz_during_balanced_double_driving}) shows that by using pulses that transform $S^z$ along geodesics in qudit sequence design, we achieve exactly the same transformation of $S^z$ as in the qubit case. This significantly simplifies the robustness condition analysis, and as we will discuss in Sec.~\ref{sec:decoupling}, allows cancelling other terms, including disorder that is proportional to $\left(S^z\right)^2$, rotation angle errors, and dominant higher-order contributions, by analogies with the qubit case. For more detailed analysis of robustness conditions, see SI Sec.~S2.(D,E).

\subsection{General Recipe for Robust Qudit Sequence Design}
\label{sec:genericrecipe}
Combining the preceding insights, we arrive at the following prescription for designing qudit robust Hamiltonian engineering sequences:
\begin{enumerate}
    \item Choose a fixed set of physically-implementable pulses; ideally ones that cause frame trajectories along geodesics.
    \item Apply the pulses a few layers deep to build a decoupling frame graph, where edges correspond to the pulses chosen above, and vertices are frames $\tilde{S}^z$ accessible using pulses in the chosen pulse set.
    \item Apply linear programming techniques described in Ref.~\cite{choi2017dynamical} to identify a subset of frames and weights that achieve decoupling.
    \item Identify a path on the decoupling frame graph that walks through all desired frames, spends the required time at each vertex, and cancels the evolution during pulses. This is usually achievable if we choose a pulse set that transforms $S^z$ along simple trajectories (e.g. geodesics) in step 1.
\end{enumerate}

The end result will be an experimentally implementable decoupling sequence that decouples both disorder and interactions, and is robust to various control imperfections.
We note that some of these conditions can be relaxed.
For example, even if the pulses do not exclusively result in geodesic precessions, we can still perform robust Hamiltonian engineering through careful design, as described in SI Sec.~S2.G.

\section{Designing a Good Qutrit Decoupling Sequence}
\label{sec:decoupling}
In this section, we will use the general recipe described in Sec.~\ref{sec:genericrecipe} to design a robust disorder and interaction decoupling sequence for a dipolar interacting spin-1 ensemble. The Hamiltonian of the system we are considering is:
\begin{align}
    H &= \sum_i \left[h_iS_i^z+D_i\left(S_i^z\right)^2\right]\nonumber\\
    &+ \sum_{ij} J_{ij}\left[S_i^z S_j^z - \frac{1}{2}H_{ij}^{XY,0+}- \frac{1}{2}H_{ij}^{XY,0-}\right],
\label{eq:spin1_Hamiltonian}
\end{align}
where the first term describes two independent modes of the on-site disorder, which we will call ``$S^z$ disorder" and ``$\left(S^z\right)^2$ disorder" from now on, and the second term is the dipole-dipole interaction after applying RWA.
The symbol $H_{ij}^{XY,0+}$ in the second term is a shorthand for the flip-flop term $H_{ij}^{XY,0+}\equiv\ket{+1,0}\bra{0,+1}+h.c.$ between $\ket{0}$ and $\ket{+1}$, and the symbol $H_{ij}^{XY,0-}$ is the similar flip-flop term between $\ket{0}$ and $\ket{-1}$.

Now let us design a robust disorder and interaction decoupling sequence using this general recipe.
We choose to work with balanced double driving pulses, because they transform $S^z$ along geodesics.
By a linear programming search on the accessible frames, which is described in more detail in Ref.~\cite{choi2017dynamical} and SI Sec.~S2.A, we find that the 12 frames in Fig.~\ref{fig:decoupling_frame_graph.png}(d) constitute a decoupling frame set when we spend equal time on each vertex.
The explicit expressions of these frames are:
\begin{align}
    \pm S^x&=\pm\frac{1}{\sqrt{2}}\left(
\begin{array}{ccc}
 0 & 1 & 0 \\
 1 & 0 & 1 \\
 0 & 1 & 0 \\
\end{array}
\right), &
\pm S^{\tilde{x}}&=\pm\frac{1}{\sqrt{2}}\left(
\begin{array}{ccc}
 0 & 1 & 0 \\
 1 & 0 & -1 \\
 0 & -1 & 0 \\
\end{array}
\right),\nonumber\\
\pm S^y&=\pm\frac{1}{\sqrt{2}}\left(
\begin{array}{ccc}
 0 & -i & 0 \\
 i & 0 & -i \\
 0 & i & 0 \\
\end{array}
\right), &
\pm S^{\tilde{y}}&=\pm\frac{1}{\sqrt{2}}\left(
\begin{array}{ccc}
 0 & -i & 0 \\
 i & 0 & i \\
 0 & -i & 0 \\
\end{array}
\right),\nonumber\\
\pm S^z&=\pm\left(
\begin{array}{ccc}
 1 & 0 & 0 \\
 0 & 0 & 0 \\
 0 & 0 & -1 \\
\end{array}
\right), &
\pm S^{\tilde{z}}&=\pm\left(
\begin{array}{ccc}
 0 & 0 & -i \\
 0 & 0 & 0 \\
 i & 0 & 0 \\
\end{array}
\right).
\label{eq:spin1_frame_set}
\end{align}

The next step is to build in robustness by choosing a good path on the decoupling frame graph. In order to build a high-performance decoupling sequence, in addition to the robustness conditions, it is also crucial for the dominant terms in the Hamiltonian to be cancelled as locally as possible to avoid generating a large higher order contribution in the Magnus expansion~\cite{magnus1954on}.
Therefore, the overall design principle involves a hierarchical structure: shorter sequences are designed to robustly cancel the dominant terms in the Hamiltonian, and they are used as building blocks for longer sequences that cancel the subdominant terms.

In our experimental platform consisting of a dense NV ensemble, the ordering of energy scales in the Hamiltonian (from large to small) is magnetic noise ($\propto S^z$), electric field noise and strain inhomogeneities ($\propto \left(S^z\right)^2$), and the dipole-dipole interaction, and we will also aim to cancel them in this order in our pulse sequence.
We note that different experimental platforms can have different relative magnitudes of these terms, and the hierarchical design method we discuss in this section should still lead to good pulse sequences in those cases.

Let us now describe the hierarchies from the lowest level to the highest level.

\begin{enumerate}
    \item \textbf{Cancel magnetic noise $S^z$}
\end{enumerate}

\begin{figure}
\begin{center}
\includegraphics[width=\columnwidth]{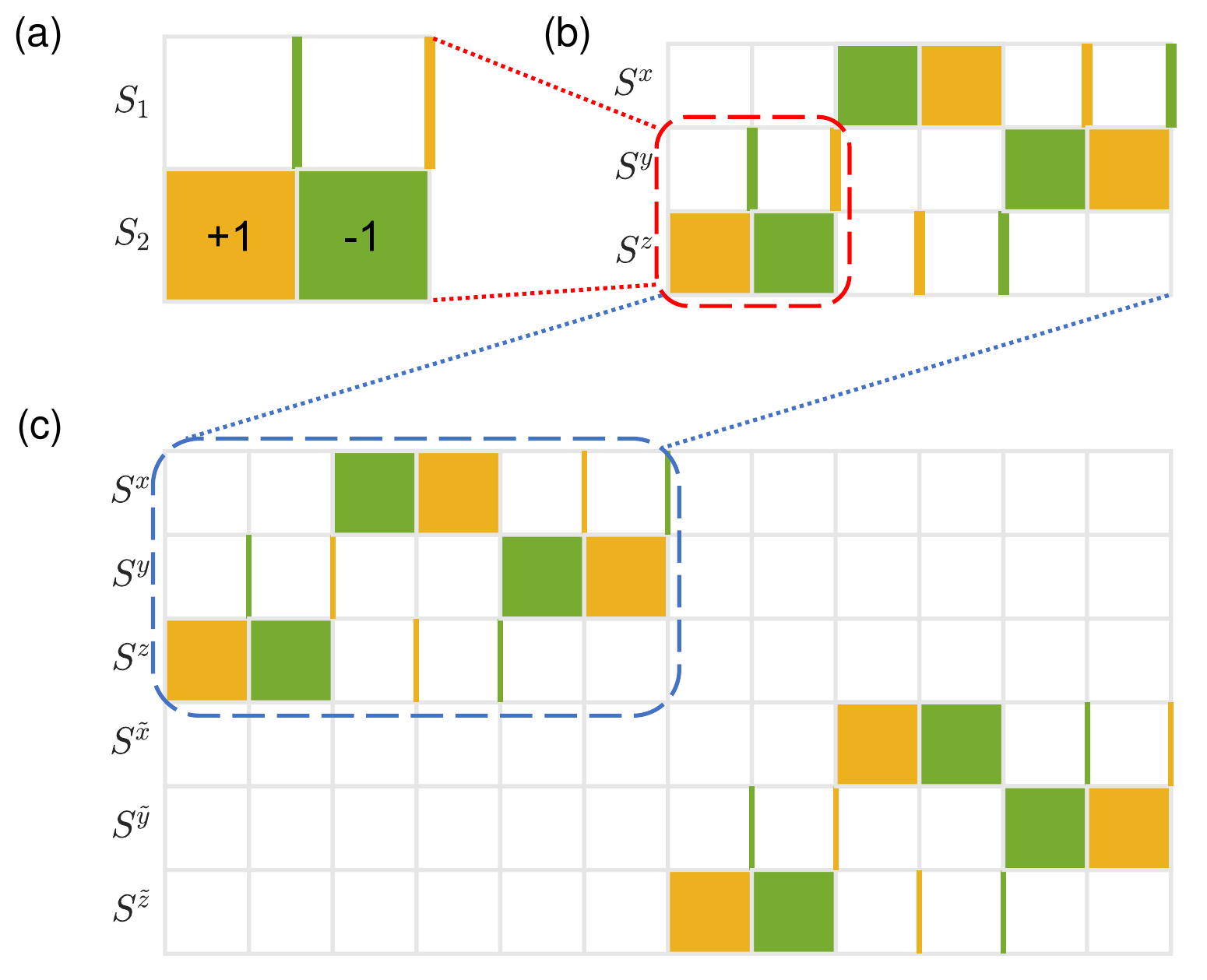}
\caption{{\bf Hierarchy of sequences.} (a) The basic building block we use to cancel $S^z$ disorder. $S_1$ and $S_2$ represent two generic frames connected by microwave pulses and yellow/green represent $+ S_{1,2}$/$-S_{1,2}$ respectively. The large square blocks indicate free evolution between pulses and the thin lines indicate intermediate frames during pulses (e.g. the thin green line in the $S_1$ row indicates a $\pi$ pulse from $S_2$ to $-S_2$ that goes through $-S_1$). Notice that disorder during both free evolution and pulses are cancelled. (b) A WAHUHA-like sequence built by the building blocks in (a) cancels the $\left(S^z\right)^2$ disorder. (c) A sequence that decouples both disorder and interactions is obtained by applying the WAHUHA sequence in (b) on each block in Fig.~\ref{fig:Two_blocks}.}
\label{fig:Hierarchy_of_sequences}
\end{center}
\end{figure}

Since magnetic noise is the dominant term in our platform, it is cancelled on the lowest level in the pulse sequence hierarchy. The basic structure to cancel the magnetic noise is shown in Fig.~\ref{fig:Hierarchy_of_sequences}(a). In these plots, we represent the pulse sequence by a frame matrix, where each column describes a frame and each row represents a basis vector we use to decompose the frames. The frames are represented by their decomposition coefficients in this basis; for example, a ``+1" in the row corresponding to $S_2$ denotes the frame $+S_2$ and a ``-1" in the $S_1$ row denotes the frame $-S_1$. The square blocks in these plots represent free evolution periods between pulses, and the thin lines represent intermediate frames we go through during pulses. For a more detailed description of these plots in the qubit case, see Ref.~\cite{choi2020robust}. In Fig.~\ref{fig:Hierarchy_of_sequences}(a), there is a pair of square blocks $+S_2$ and $-S_2$, so the $S^z$ disorder during free evolution is cancelled; there is also a pair of thin lines $-S_1$ and $+S_1$, so the $S^z$ disorder during pulses is also cancelled. Therefore, this basic structure cancels $S^z$ disorder robustly, and we will use it as the building block for higher level sequences that cancel other terms in the Hamiltonian.

\begin{enumerate}
    \item [2.] \textbf{Cancel electric field and strain noise $(S^z)^2$}
\end{enumerate}

\begin{figure}
\begin{center}
\includegraphics[width=\columnwidth]{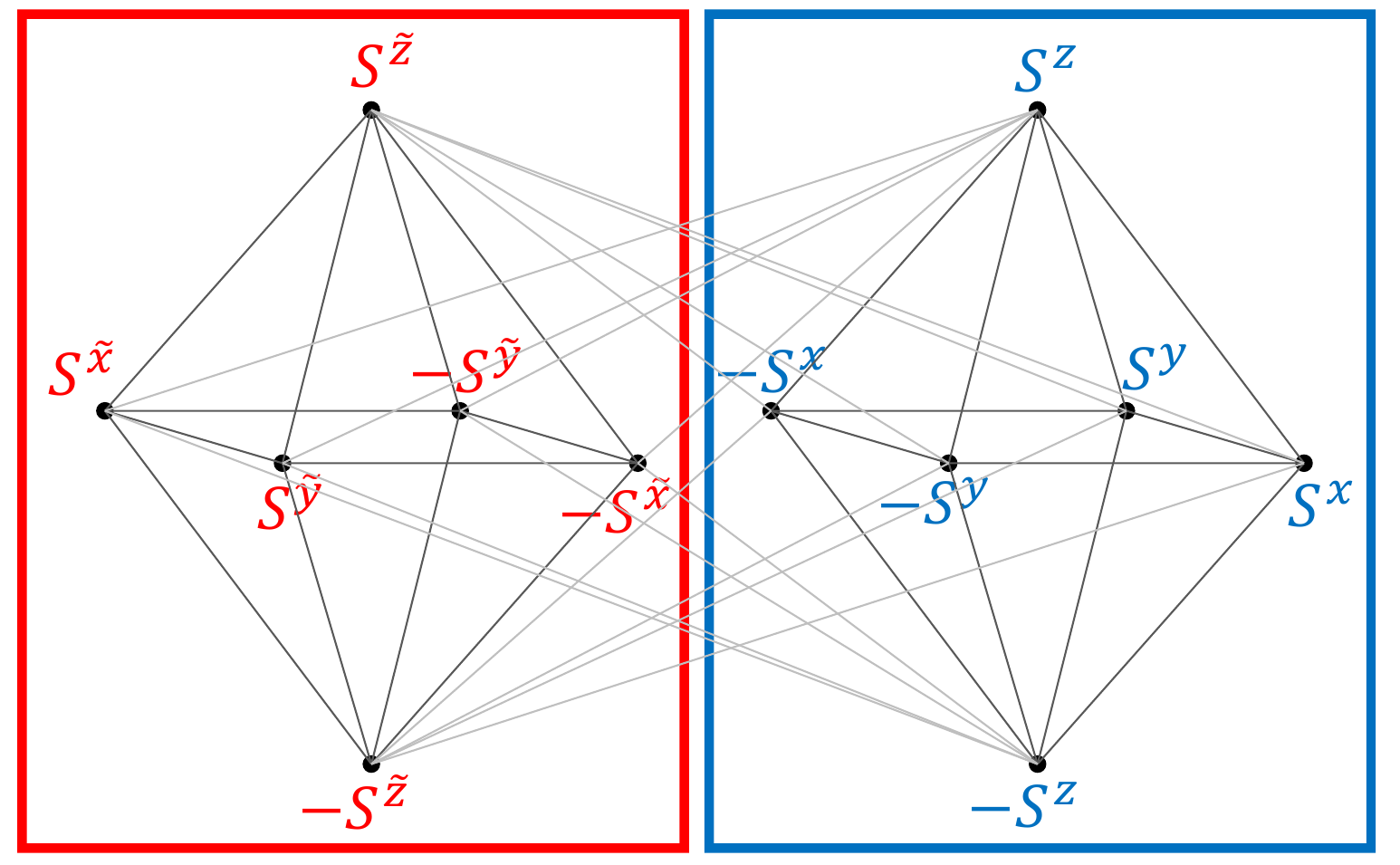}
\caption{{\bf Dividing the 12 frames into two pseudo Bloch spheres.} The 6 frames in each pseudo Bloch sphere ($\pm\hat{S}_1, \pm\hat{S}_2, \pm\hat{S}_3$) satisfy $\hat{S}_1^2+\hat{S}_2^2+\hat{S}_3^2\propto\mathbb{1}$; therefore, a WAHUHA sequence in each pseudo Bloch sphere cancels the $\left(S^z\right)^2$ disorder.
Note that additional connections from $\pm S^x$ to $\pm S^{\tilde{x}}$ and from $\pm S^y$ to $\pm S^{\tilde{y}}$ are not drawn for visual clarity.
The illustrated way of dividing the 12 frames into two pseudo-Bloch-spheres that satisfy the requirement $\hat{S}_1^2+\hat{S}_2^2+\hat{S}_3^2\propto\mathbb{1}$ is not unique.
In fact, there are 4 such divisions and picking other divisions will lead to similar pulse sequences.}
\label{fig:Two_blocks}
\end{center}
\end{figure}

The next level in the hierarchy is cancelling electric field noise and strain inhomogeneities. An important observation here is that the WAHUHA sequence~\cite{waugh1968approach}, which is designed to cancel spin-$\frac{1}{2}$ XXZ interactions, can also cancel the spin-1 $\left(S^z\right)^2$ disorder. Specifically, consider a spin-1 version of the WAHUHA sequence that goes through the frames $S^x$, $S^y$, and $S^z$, where $S^{x,y,z}$ are conventional spin-1 operators (see above). The reason that this sequence cancels the $\left(S^z\right)^2$ disorder is because it transforms the disorder into $\left(\hat{S}^x\right)^2+\left(\hat{S}^y\right)^2+\left(\hat{S}^z\right)^2=\hat{S}^2=S(S+1)\mathbb{1}\propto\mathbb{1}$, which is a trivial constant. Moreover, due to the similar structure of $\left(\hat{S}^z\right)^2$ and the spin-$\frac{1}{2}$ interaction $\hat{S}^z\otimes\hat{S}^z$, both being quadratic in $S^z$, their contribution during the finite pulse can be cancelled using the same method, as discussed in more detail in the fourth level of the hierarchy and in Ref.~\cite{choi2020robust}.
In addition, we find that the 12 frames in Fig.~\ref{fig:decoupling_frame_graph.png}(d) can be divided into 2 blocks of 6 frames (as shown in Fig.~\ref{fig:Two_blocks}), in which the frames in each block ($\pm\hat{S}_1, \pm\hat{S}_2, \pm\hat{S}_3$) satisfy $\hat{S}_1^2+\hat{S}_2^2+\hat{S}_3^2\propto\mathbb{1}$, achieving the same cancellation as above. Therefore, we can cancel the $\left(S^z\right)^2$ disorder locally by applying a WAHUHA sequence on each block, as shown in Fig.~\ref{fig:Hierarchy_of_sequences}(b). Note however that a WAHUHA sequence does not fully cancel the $\left(S^z\right)^2$ disorder during the finite pulse duration, and we postpone this cancellation to the fourth level.

\begin{enumerate}
    \item [3.] \textbf{Cancel dipole-dipole interactions}
\end{enumerate}

The third level in the hierarchy is to cancel the dipole-dipole interaction. As we found from our linear programming analysis in Sec.~\ref{sec:graphical}, this requires us to spend equal time in all 12 frames in Fig.~\ref{fig:Two_blocks} and is achieved by concatenating the WAHUHA sequences on the two blocks in Fig.~\ref{fig:Two_blocks}. The frame representation of this sequence is shown in Fig.~\ref{fig:Hierarchy_of_sequences}(c).

\begin{enumerate}
    \item [4.] \textbf{Further improvements inspired by advanced qubit sequence design}
\end{enumerate}

Inspired by recent advances in qubit higher-order sequence design \cite{zhou2023robust} and the similarity between certain terms in qutrit decoupling and qubit decoupling, we can use the interaction decoupling sequence shown in Fig.~\ref{fig:Hierarchy_of_sequences}(c) as a building block, and apply the higher order designs in Ref.~\cite{zhou2023robust} to further improve its performance.
The essence of these further improvements is that we are flipping the signs and ordering of the frames to cancel $\left(S^z\right)^2$ disorder during the pulses, the Rabi inhomogeneity effect, and the first order terms in the Magnus expansion coming from commutators between various terms. 

As an example, the $\left(S^z\right)^2$ disorder during the pulses and the Rabi inhomogeneity effect are cancelled by flipping the signs of the intermediate frames (or the free evolution frames, but not both) in the basic building block shown in Fig.~\ref{fig:Hierarchy_of_sequences}(a).
The cancellation of $\left(S^z\right)^2$ disorder during pulses relies on the fact that it transforms in the same way as spin-$\frac{1}{2}$ Ising interactions. Specifically, if we denote the frame before and after a spin-1 $\frac{\pi}{2}$ pulse by $S_1$ and $S_2$, and denote the angle rotated from $S_1$ by $\theta$, then the instantaneous frame is $\cos{\theta}S_1+\sin{\theta}S_2$ (see Sec.~\ref{sec:robusthigherspin} for the derivation). The $\left(S^z\right)^2$ disorder is thus transformed to $\cos^2{\theta}S_1^2+\sin^2{\theta}S_2^2+\sin{\theta}\cos{\theta}(S_1S_2+S_2S_1)$. The terms proportional to $S_1^2$ and $S_2^2$ can be viewed as additional time spent in the frame right before and right after, and therefore get cancelled by the WAHUHA block itself.
The cross term $S_1S_2+S_2S_1$ is cancelled here because one of $S_1$ and $S_2$ is an intermediate frame whose sign is flipped, in analogy to the rule for interaction cross-terms in Ref.~\cite{choi2020robust}.
Meanwhile, the Rabi inhomogeneity effect is cancelled because the rotation direction changes when the sign of one frame in $S_1$ and $S_2$ is flipped, leading to forward and backward rotations that compensate each other.
This is in direct analogy to the chirality sum rule in Ref.~\cite{choi2020robust}. 
Furthermore, by similar analysis, one can show that the final sequence is not only robust to rotation angle errors common to both transitions as discussed above, but also robust to rotation angle errors on each individual transition (see SI Sec.~S2.E for details).
Similar analogies to the qubit case also apply to higher-order contributions.

With the preceding hierarchical construction, we arrive at a set of promising decoupling pulse sequences, as described in full detail in the SI Sec.~S2.B. For applications on other experimental platforms, the ordering of level 1 to 3 in the hierarchy can be changed based on the relative magnitude of disorder and interactions; the symmetrizations in level 4 are optional based on the trade off between better decoupling performance versus shorter sequence length.

\section{Qutrit Decoupling Experiment}
\label{sec:experiment}
\begin{figure}
\begin{center}
\includegraphics[width=\columnwidth]{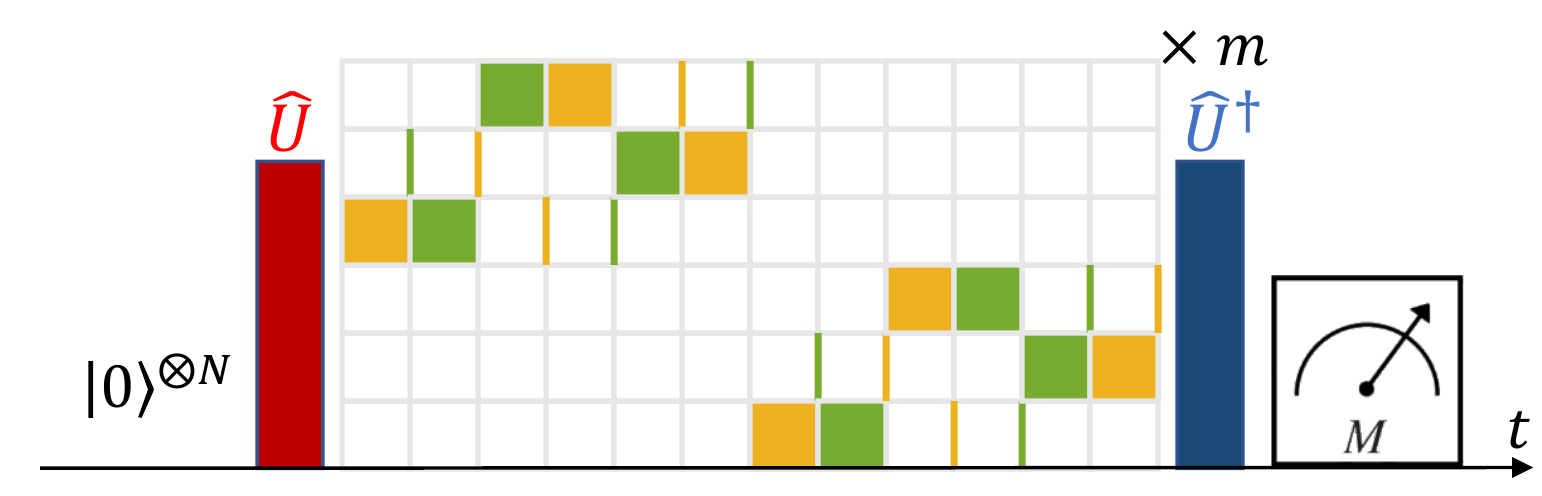}
\caption{{\bf The decoupling experiment.} In the experiment, we first initialize the NVs to state $\ket{0}$ by shining a $532$ nm green laser to our sample, then apply the initialization microwave pulse $\hat{U}$ to prepare the initial state whose decay curve we want to measure. After that, the decoupling pulse sequence is applied, and finally the preparation pulse is reversed before measuring the population in state $\ket{0}$.}
\label{fig:Decoupling_experiment}
\end{center}
\end{figure}

We now test the performance of the robust qutrit decoupling sequence proposed in Sec.~\ref{sec:decoupling} in a high density ensemble of spin-1 NV centers in diamond~\cite{kucsko2018critical,zhou2020quantum}, resulting in the first demonstration of full decoupling of qudit interactions. We isolate NVs with the same lattice orientation with an external magnetic field aligned with one group of NV centers. This magnetic field also breaks the degeneracy between energy levels $\ket{\pm 1}$, allowing us to address the transitions $\ket{0}\leftrightarrow\ket{+1}$ and $\ket{0}\leftrightarrow\ket{-1}$ separately, using microwave with frequencies $3.647~\text{GHz}$ and $2.092~\text{GHz}$, respectively. The density of NV centers along each lattice orientation in our sample is about 4~ppm, which corresponds to a typical interaction strength of $J=2\pi\times 35~\text{kHz}$. The strength of the onsite $S^z$ disorder and $\left(S^z\right)^2$ disorder is about $2\pi\times 4~\text{MHz}$ and $2\pi\times 1~\text{MHz}$ respectively (Gaussian standard deviation).
In the experiment, we optically initialize the state of NVs to be in $\ket{0}$, apply microwave pulses to prepare various initial states, then apply the decoupling sequence, and finally reverse the preparation sequence before reading out the population in state $\ket{0}$ via fluorescence differences (see Fig.~\ref{fig:Decoupling_experiment} and Ref.~\cite{kucsko2018critical} for more details).

\begin{figure}
\begin{center}
\includegraphics[width=\columnwidth]{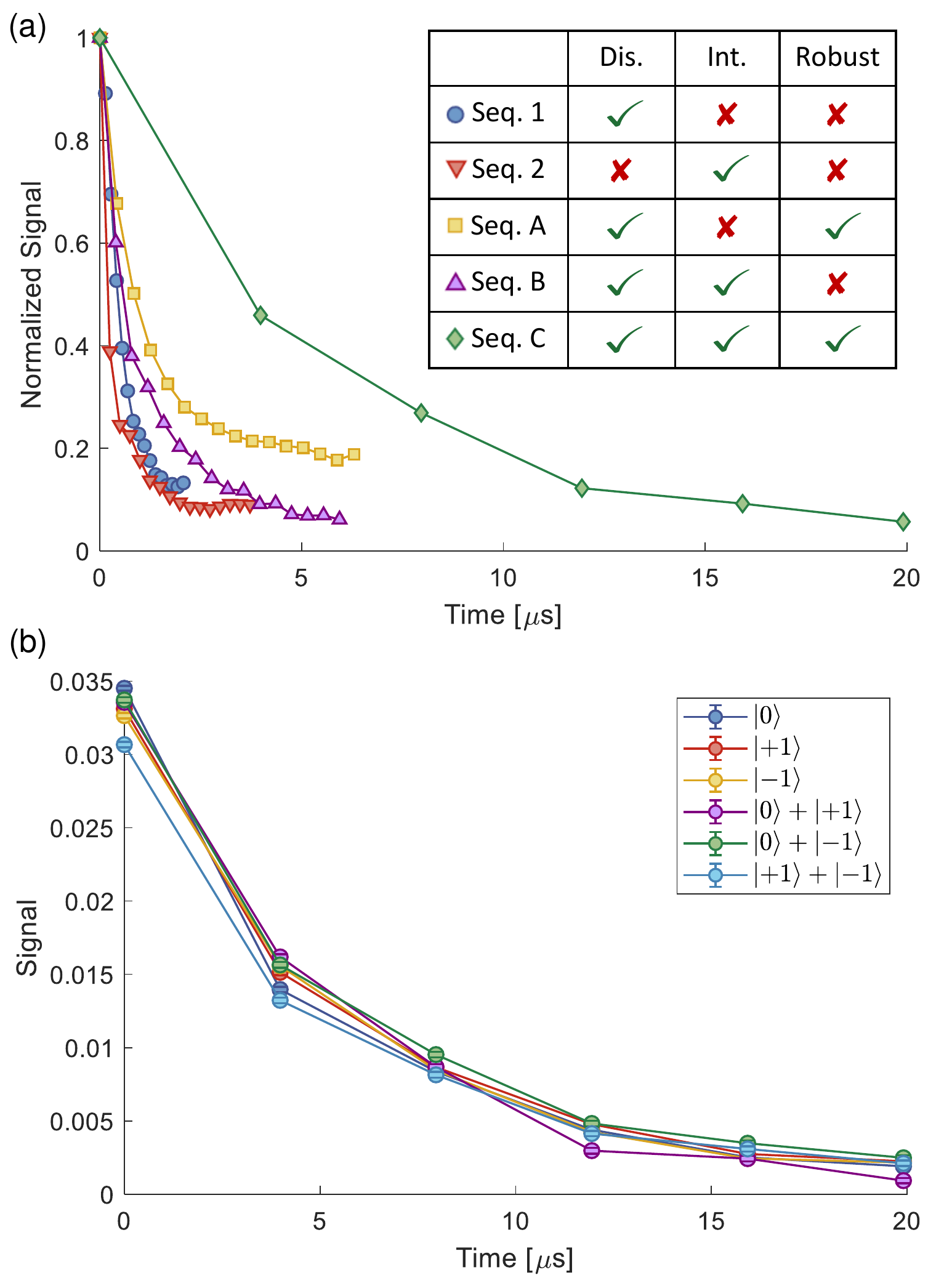}
\caption{{\bf Decoupling experiment results.} (a) Measured average decay trace for different pulse sequences, where the average is taken over all three coherent superposition initial states $\frac{\ket{0}+\ket{-1}}{\sqrt{2}}$, $\frac{\ket{0}+\ket{+1}}{\sqrt{2}}$, and $\frac{\ket{+1}+\ket{-1}}{\sqrt{2}}$. Among the sequences, Seq.~1,2 are existing sequences in Ref.~\cite{choi2017observation} and \cite{choi2017dynamical}, respectively, while Seq.~A,B,C are new sequences we designed.
The inset table shows whether or not a sequence decouples disorder, interaction, and achieves robustness against disorder during pulses.
For more details on these sequences, see SI Sec.~S2.(B,C).
The measurement is performed with a differential readout, where we rotate the population in each of the three states $\ket{+1}$, $\ket{0}$, and $\ket{-1}$ to state $\ket{0}$ before doing the fluorescence measurement.
Denoting the measured fluorescence by $I_+$, $I_0$, and $I_-$ respectively, the signal on the vertical axis is defined as $S=\frac{3}{2}\frac{2I_0-I_--I_+}{I_0+I_-+I_+}$, which is proportional to $P_0-\frac{1}{3}$, where $P_0$ is the population in state $\ket{0}$.
The experimental parameters are $\left(t_{\pi/2}=8~\text{ns}, \tau=25~\text{ns}\right)$, where $t_{\pi/2}$ is the time duration of each spin-1 $\frac{\pi}{2}$ pulse and $\tau$ is the time spent in each frame. (b) Decay curve of different initial states for our best decoupling sequence DROID-C3PO (Seq.~C), showing that the sequence preserves arbitrary initial quantum state.} 
\label{fig:Decoupling_experiment_results}
\end{center}
\end{figure}

The measured decay of the signal under various decoupling sequences is plotted in Fig.~\ref{fig:Decoupling_experiment_results}(a).
Sequences with numerical labels are existing sequences from the literature, while the ones with alphabetical labels are new sequences we designed.
Seq.~2 is the interaction decoupling sequence in Ref.~\cite{choi2017dynamical}. Its performance in our experiment is not good, because it does not decouple the disorder, which is the dominant term in our system. To cancel the disorder, we can use Seq.~1 from Ref.~\cite{choi2017observation} that directly generalizes the spin-$\frac{1}{2}$ spin-echo sequence to the spin-1 case. Similar to the spin-echo sequence, Seq.~1 only cancels disorder during the free evolution, but is not robust to disorder during pulses.

To improve the performance, we designed Seq.~A, which is an enhanced version of Seq.~1 that is highly robust to disorder effects during pulses. This robust disorder decoupling sequence shows a significant timescale extension compared to its non-robust counterpart Seq.~1, highlighting the importance of robust sequence design. Furthermore, since Seq.~A does not cancel interactions, it serves as a baseline for verifying interaction decoupling in further sequences. To decouple both disorder and interaction, we designed Seq.~B, which is the sequence shown in Fig.~\ref{fig:decoupling_frame_graph.png}(e,f). Although this sequence further decouples interaction, its performance in experiment is worse than Seq.~A, because it has no robustness built in. For more detailed description of these sequences, see SI Sec.~S2.(B,C).

Most importantly, after integrating all robustness considerations into the sequence design, we arrive at our current best sequence Seq.~C, which we call ``DROID-C3PO" (i.e. Disorder-RObust Interaction Decoupling - Coherent 3-level Pulse Optimization). This sequence decouples both disorder and interactions, and is robust to disorder during pulses, rotation angle errors, and dominant higher order contributions. In the experiment, this sequence shows significant improvement over Seq.~A, constituting the first demonstration of full disorder and interaction decoupling in a qudit system, and achieves a ten-fold improvement over the existing sequences. In addition, we verify in Fig.~\ref{fig:Decoupling_experiment_results}(b) that the decay timescales of all initial states are extended under Seq.~C, confirming that this sequence is a true decoupling sequence that preserves an arbitrary quantum state. The complete plot of Seq.~C is shown in Fig.~\ref{fig:Main_Results}(f), and its frame matrix representation is shown in Fig.~S3(a).

\section{Many-Body Physics: Quantum Many-Body Scars}
\label{sec:scars}

The same techniques developed above can also be used to engineer a rich family of interesting many-body Hamiltonians, which enables new phenomena not accessible in spin-$\frac{1}{2}$ systems.
As a specific example, we will discuss the engineering of a Hamiltonian that supports quantum many-body scars---exotic non-thermalizing eigenstates embedded in an otherwise thermal spectra, which constitute a new class of thermalization phenomena in between thermalizing systems and many-body localized systems~\cite{bernien2017probing,bluvstein2021controlling,kao2021topological,turner2018weak,ho2019periodic,lin2019exact,khemani2019signatures,choi2019emergent,maskara2021discrete,schecter2019weak}.

A recent paper~\cite{schecter2019weak} proposed that the bipartite spin-1 XY model naturally realizes quantum many-body scars.
Specifically, the model contains two groups of spin-1 particles with no intragroup interactions but with intergroup $XX+YY$ interactions. The Hamiltonian for this model is given by
\begin{equation}
    H = \sum_{i\in A, j\in B} J_{ij}\left(S_i^x S_j^x + S_i^y S_j^y\right) + h\sum_{i\in A, B} S_i^z,
\label{eq:Scar_hamiltonian}
\end{equation}
where spins $i$ and $j$ reside in different groups $A$ and $B$ in the first term, and $h$ is an external magnetic field coupled to $S^z$.

In this particular example, the scar subspace is formed by acting with raising operators which act only on the $\ket{+1},\ket{-1}$ subspace. More concretely, for the bipartite spin-1 XY model we are considering, we can define the $SU(2)$ algebra operators:
\begin{align}
    J^\pm &= \frac{1}{2}\sum_i a_i\left(S_i^\pm\right)^2, \quad J^z= \frac{1}{2}\sum_i S_i^z,
\end{align}
where $a_i=1/-1$ for spins in group A/B, and $S_i^\pm, S_i^z$ are spin-1 raising, lowering, and $S^z$ operators.
With this notation, the following eigenstates
\begin{align}
\ket{\mathcal{S}_n}\propto (J^+)^n\ket{-1}^N
\label{eq:scar_subspace}
\end{align}
form the non-thermalizing scar manifold according to Ref.~\cite{schecter2019weak}, where $\ket{-1}^N$ is the state with all spins fully polarized into $\ket{-1}$, and $N$ is the total number of spins in the two groups.
At the same time, the Hamiltonian itself does not commute with the subspace $SU(2)$ generator $J^\pm$, indicating that it is not integrable.
Indeed, one can verify that this Hamiltonian has a thermal spectrum~\cite{schecter2019weak}, where generic initial states thermalize.
Thus, the spin-1 XY model constitutes a quantum many-body scar.

\begin{figure}
\begin{center}
\includegraphics[width=\columnwidth]{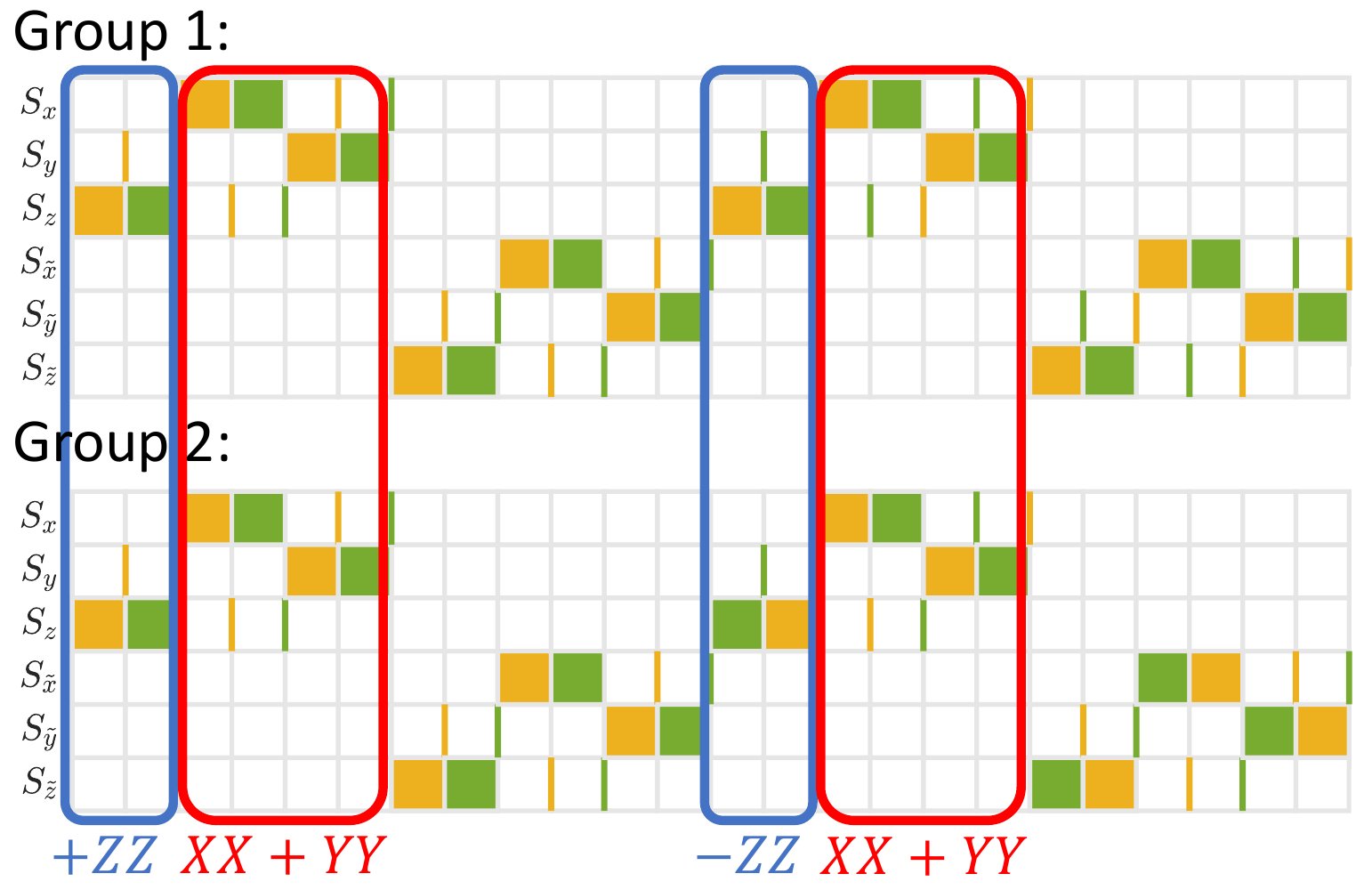}
\caption{{\bf The sequence engineering the scar Hamiltonian.} The whole sequence to engineer the scar Hamiltonian $H = \sum_{i\in A, j\in B} J_{ij}\left(S_i^x S_j^x + S_i^y S_j^y\right)$ is plotted. Note that all intragroup interactions are decoupled by applying a decoupling sequence on each group individually, and that because of the sign flip in the second half of the sequence for group 2, all intergroup interactions are cancelled (see blue boxes) except the $S^xS^x$ and $S^yS^y$ terms, which add up (see red boxes) and give the desired bipartite spin-1 XY model.}
\label{fig:Scar_hamiltonian_engineering}
\end{center}
\end{figure}

Using our techniques, the bipartite spin-1 XY model discussed above can be engineered from the native dipole-dipole interaction in high density NV center samples.
Here, the two groups in the model can be realized as NV centers along two lattice orientations, where the transition frequencies of the two groups are spectrally resolved and the two groups can be controlled independently.
The intragroup interaction can be cancelled by applying the robust interaction decoupling sequence in each group as discussed above.
To engineer the intergroup XY interaction, notice that the two groups of NV centers along different lattice orientations are not on resonance with each other when an external magnetic field is applied.
Therefore, the interaction between NVs residing in different groups is an Ising interaction $S^z\otimes S^z$.
A simple way to engineer the intergroup XY Hamiltonian is thus to repeat the basic sequence twice, and in the second iteration flip the signs of the frame pairs \{$\pm S^z, \pm S^{x^\prime}, \pm S^{y^\prime}, \pm S^{z^\prime}$\} on the second group of NVs while leaving the signs of the frame pairs \{$\pm S^x, \pm S^y$\} unchanged (see Fig.~\ref{fig:Scar_hamiltonian_engineering}). In this way, the $-S^zS^z, -S^{x^\prime}S^{x^\prime}, -S^{y^\prime}S^{y^\prime}, -S^{z^\prime}S^{z^\prime}$ interactions in the second iteration cancel with the $+S^zS^z, +S^{x^\prime}S^{x^\prime}, +S^{y^\prime}S^{y^\prime}, +S^{z^\prime}S^{z^\prime}$ interactions in the first iteration; while the $+S^xS^x, +S^yS^y$ interactions in both iterations add up and gives the desired XY Hamiltonian.
A frame representation of the pulse sequence that engineers the scar Hamiltonian is shown in Fig.~\ref{fig:Scar_hamiltonian_engineering}.

\begin{figure}
\begin{center}
\includegraphics[width=\columnwidth]{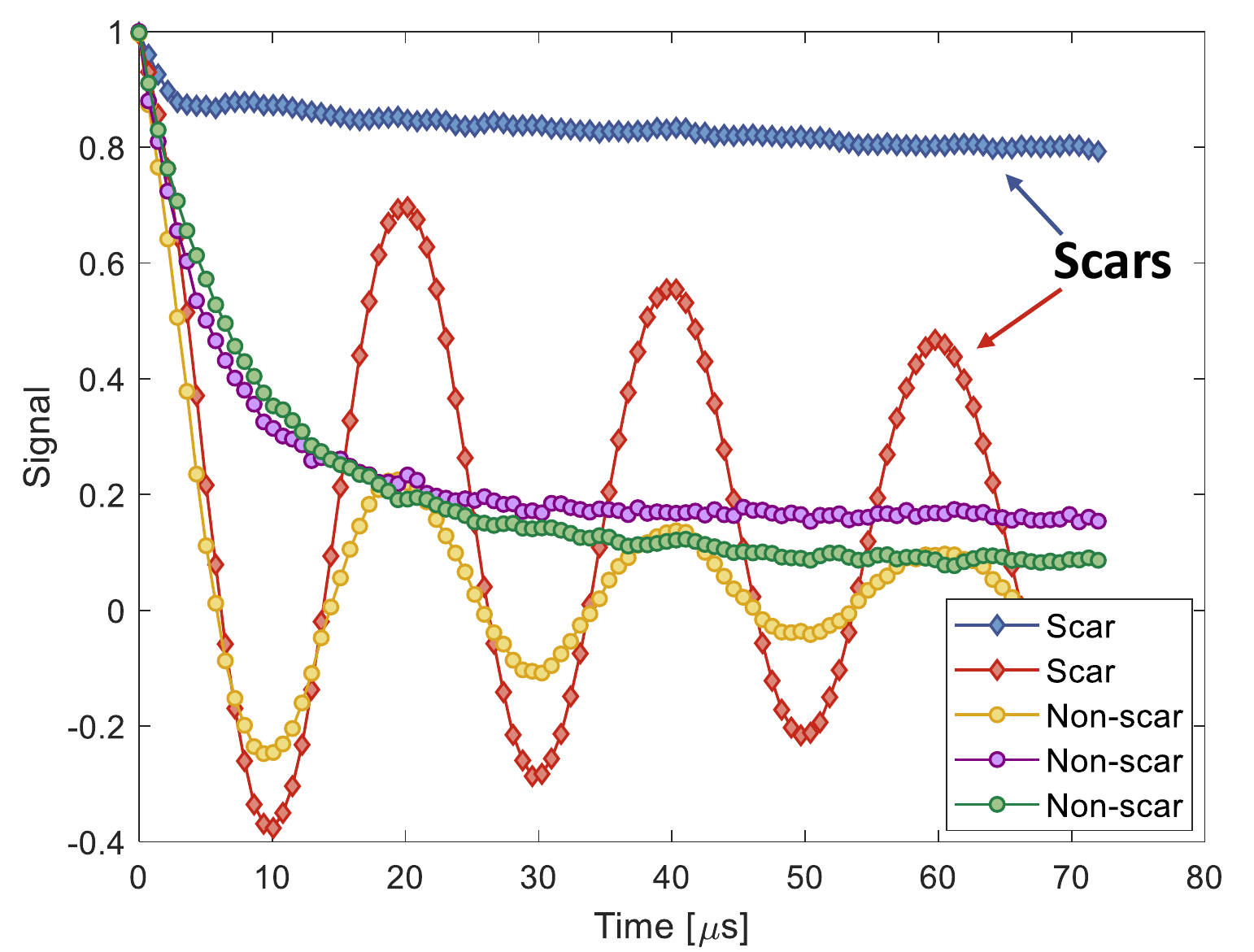}
\caption{{\bf Numerical simulation of the scar dynamics under the sequence in Fig.~\ref{fig:Scar_hamiltonian_engineering}.} We simulated the dynamics of various initial states under the sequence that engineers the scar Hamiltonian (Eq.~(\ref{eq:Scar_hamiltonian})). The experiment we simulated is the same type of experiment (i.e. initialize, evolve, readout) as in Fig.~\ref{fig:Decoupling_experiment}. The simulation parameters are $(\Delta=0, J=2\pi\times 35~\text{kHz}, h=2\pi\times 25~\text{kHz}, t_{\pi/2}=5~\text{ns},$ $\tau=20~\text{ns}, N_1=N_2=4)$, where $\Delta$ is the disorder strength, $t_{\pi/2}$ is the time duration of each $\frac{\pi}{2}$ pulse, $\tau$ is the free evolution time spent in each frame, and $N_1, N_2$ are the number of spins in the two groups.}
\label{fig:Scar_simulation}
\end{center}
\end{figure}

We simulated the dynamics of various initial states under this pulse sequence. The simulated initial states include $\left(\frac{\ket{+1}+\ket{-1}}{\sqrt{2}},\frac{\ket{+1}-\ket{-1}}{\sqrt{2}}\right)$, $\left(\ket{+1},\ket{+1}\right)$, $\left(\frac{\ket{+1}+\ket{-1}}{\sqrt{2}},\frac{\ket{+1}+\ket{-1}}{\sqrt{2}}\right)$, $\left(\ket{+1},\ket{-1}\right)$, $\left(\ket{0},\ket{0}\right)$, where the first state in the bracket represents the initial state of the first group of spins and the second state in the bracket represents the initial state of the second group of spins. Based on the geometric intuition discussed in SI Sec.~S2.F (which states that the scar subspace $\ket{S_n}$ is the maximal spin subspace after rotating the second group of spins by $\pi$ around the $z$ axis), the first two states live in the scar subspace and therefore are not expected to thermalize, while the last three states do not live in the scar subspace and are expected to thermalize. The simulated dynamics of these initial states is plotted in Fig.~\ref{fig:Scar_simulation}. In the plot, we see that the initial states living in the scar subspace do not thermalize (their signal either stays large or has persistent oscillations); while the signals for other initial states quickly decay away. These results show that exotic quantum many-body scar states can be observed even in highly disordered, natural systems such as randomly positions ensembles of NVs, in contrast to the more regular, clean systems in which they have been observed to date~\cite{bernien2017probing,bluvstein2021controlling,kao2021topological}.

\section{Enhanced Quantum Sensing with Qudit Hamiltonian Engineering}
\label{sec:sensing}
In addition to a rich landscape of many-body Hamiltonians, higher spin systems also give rise to interesting opportunities in quantum sensing.
First, the higher spin implies a larger effective dipolar moment, which can lead to a linear or quadratic enhancement in magnetic field sensitivity, depending on the nature of the signal~\cite{myers2017double}.
A well-known example of this for non-interacting spins is the use of double-quantum magnetometry for nitrogen-vacancy centers~\cite{fang2013high,mamin2014multipulse,bauch2018ultralong,hart2021nv}, and interacting spin systems present further challenges and opportunities for sensing sequence design~\cite{zhou2020quantum,balasubramanian2019dc}.
Second, the larger Hamiltonian design space may also enable full time reversal of the interaction Hamiltonian, useful for entanglement-enhanced sensing~\cite{goldstein2011environment,davis2016approaching}, which may not be otherwise accessible with a subset of levels~\cite{kucsko2018critical}.
For example, the spin-1 dipolar interaction Hamiltonian projected onto $|0\rangle$ and $|+1\rangle$ has a nonzero trace when expressed in the Pauli basis, resulting in a Heisenberg interaction component that cannot be reversed through global drives; but the full spin-1 dipolar interaction can nevertheless be fully cancelled.
In this section, we will provide a systematic understanding of how to evaluate the sensitivity for a given sensing sequence, which is determined by the difference between the largest and smallest eigenvalue of the transformed toggling frame operator, and provide simple examples to illustrate this.
We leave the systematic design of sensing-oriented pulse sequences for higher spin systems to future work.

In order to perform quantum sensing, we add to the Hamiltonian a term corresponding to the target sensing field:
\begin{align}
    H=H_{\textrm{dis}}+H_{\textrm{int}}+H_{\textrm{sense}}.
\end{align}
We focus on the case of sensing time-dependent magnetic fields, where $H_{\textrm{sense}}(t)=\sum_i B(t)S_i^z$, although the same techniques can be readily adapted to electric field or strain sensing, among others.
Note that the rotating wave approximation implies that $H_{\textrm{sense}}$ will be diagonal, regardless of the type of target sensing field, and can thus be written as a polynomial in $S_i^z$.

Under the Hamiltonian engineering transformations, the sensing Hamiltonian will be transformed accordingly, and the effective average Hamiltonian contribution becomes
\begin{align}
    \bar{H}_{\textrm{sense}}=\frac{1}{T}\int_0^T dt B(t)\tilde{S}_i^z(t).
\end{align}
This can be readily evaluated based on the instantaneous toggling frame transformations $\tilde{S}^z(t)$.

\begin{figure}
\begin{center}
\includegraphics[width=\columnwidth]{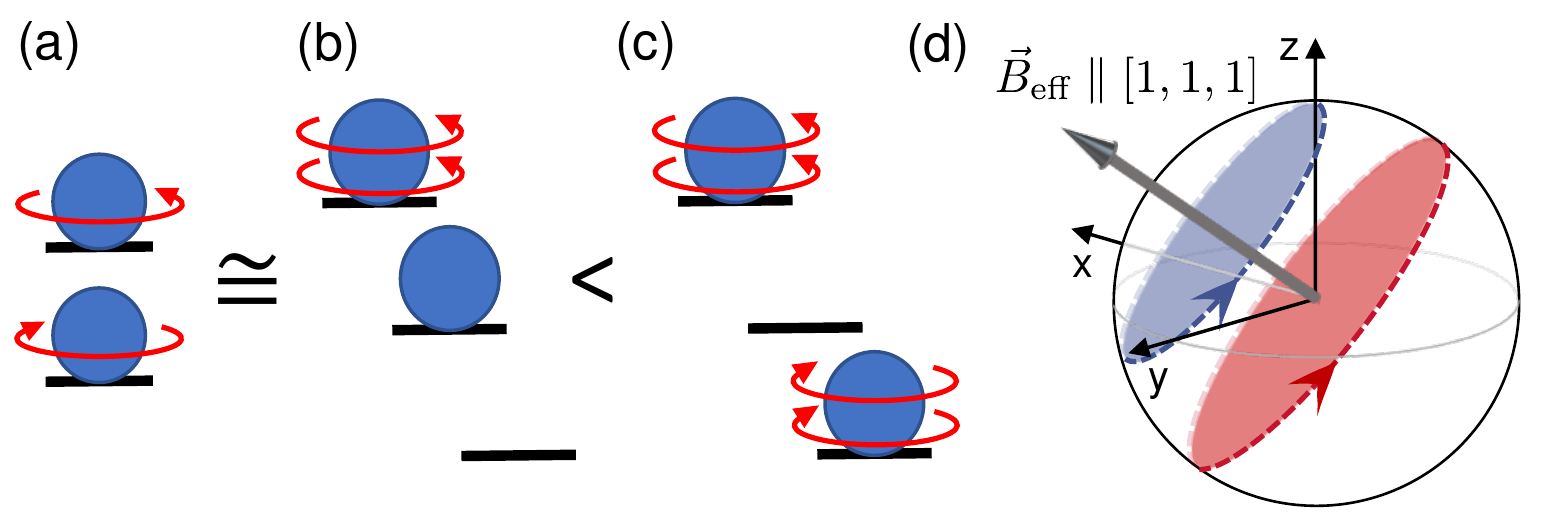}
\caption{{\bf Sensing with higher spin.} (a) Sensing with spin-$\frac{1}{2}$ particles involves preparing spins in an equal superposition of the two basis states.
Preparing the same initial state in a spin-1 system (b) yields a worse sensitivity than preparing a superposition of maximal and minimal eigenvalues in the spin-1 system (c). Number of arrows in (a-c) represents the phase accumulation speed.
(d) Preparing the initial state to be a superposition of the largest and smallest eigenvalues maximizes precession.
For the spin-$\frac{1}{2}$ case, this is achieved by preparing an initial state that is perpendicular to the effective field.}
\label{fig:sensing}
\end{center}
\end{figure}

The sensitivity to an external magnetic field is generally characterized by the quantum Fisher information (QFI)~\cite{liu2019quantum,ma2011quantum}.
In our case, since the sensing field only involves single body operators, we can directly read off the optimal initial state and measurement axis that maximizes the QFI;
we simply prepare an equal superposition between the eigenstates of $\bar{H}_{\textrm{sense}}$ with the largest and smallest eigenvalues, and do a Ramsey experiment within this two-level subspace.
This will maximize the amount of phase accumulation under a weak perturbation, achieving the best possible magnetic field sensitivity for a given pulse sequence.

Let us illustrate this with a few concrete examples.
First, consider the case of quantum sensing with interacting spin-$\frac{1}{2}$ spin ensembles.
As described in Ref.~\cite{zhou2020quantum,waugh1968approach}, the conditions for interaction decoupling transform the original target sensing field $BS^z$ into an effective sensing field $B(S^x+S^y+S^z)/\sqrt{3}$.
The largest and smallest eigenvectors are spin states aligned and anti-aligned with the sensing field direction, and the optimal initial state will be a spin state prepared in the plane orthogonal to the sensing field.
This maximizes the precession around the sensing field, as illustrated in Fig.~\ref{fig:sensing}(d), consistent with the results of Ref.~\cite{zhou2020quantum}.

We can also use the same technique to calculate the sensitivity of existing pulse sequences in the literature to a DC magnetic field.
For simplicity, we assume ideal, infinitely fast pulses, and consider the average Hamiltonian contribution from a DC magnetic field for both spin-1 pulse sequences considered in Ref.~\cite{choi2017dynamical} and Ref.~\cite{okeeffe2019hamiltonian}, as well as the famous spin-$\frac{1}{2}$ WAHUHA sequence in Ref.~\cite{waugh1968approach}.
The average Hamiltonian $\bar{H}_{\textrm{sense}}$ and its eigenvalues $\lambda$ are summarized in Table.~\ref{tab:sensitivity_comparison}, where we see a larger eigenvalue difference $\Delta\lambda$ for the two spin-1 sequences compared to the spin-$\frac{1}{2}$ sequence, indicating a higher spin enhanced sensitivity. We also find that contrary to the suggestion in Ref.~\cite{okeeffe2019hamiltonian} that HoRD-qutrit-8 is better for sensing, the sequence CYL-6 in Ref.~\cite{choi2017dynamical} has a larger eigenvalue difference, implying a higher sensitivity when preparing the optimal initial state.
This highlights the importance of evaluating sensitivity using our approach of examining eigenvalue differences.

\begin{table}[]
    \centering
    \begin{tabular}{|c|c|c|c|}
        \hline
         Sequence Name & $\bar{H}_{\textrm{sense}}$ & $\lambda$ & $\Delta\lambda$  \\\hline

         WAHUHA\cite{waugh1968approach} & $\left(\begin{array}{cc}  \frac{1}{6} & \frac{1-i}{6}\\  \frac{1+i}{6} & -\frac{1}{6} \\ \end{array}\right)$ & $\begin{array}{c} -0.289\\ ~~0.289 \end{array}$ & 0.577  \\\hline

         CYL-6\cite{choi2017dynamical}& $\left(\begin{array}{ccc}
 \frac{1}{6} & \frac{\sqrt{2}i}{6} & -\frac{i}{6} \\
 -\frac{\sqrt{2}i}{6} & 0 & -\frac{\sqrt{2}}{6} \\
 \frac{i}{6} & -\frac{\sqrt{2}}{6} & -\frac{1}{6} \\
\end{array}\right)$ & $\begin{array}{c} -0.333\\ -0.122\\ ~~0.455 \end{array}$ & 0.789  \\\hline

         HoRD-qutrit-8\cite{okeeffe2019hamiltonian} & $\left(\begin{array}{ccc}
 \frac{1}{3} & 0 & 0 \\
 0 & 0 & 0 \\
 0 & 0 & -\frac{1}{3} \\
\end{array}\right)$ & $\begin{array}{c} -0.333\\ ~~0\\ ~~0.333 \end{array}$ & 0.667  \\\hline
    \end{tabular}
    \caption{{\bf $\bar{H}_{\textrm{sense}}$ and $\lambda$ for three sensing sequences.} We see a larger $\Delta\lambda$ for the two spin-1 sequences compared to the spin-$\frac{1}{2}$ sequence, indicating a higher-spin-enhanced sensitivity. By further comparing the two spin-1 sequences, we see that their eigenvalue differences $\Delta\lambda$ are not the same, and CYL-6 has a larger eigenvalue difference despite having smaller diagonal matrix elements.}
    \label{tab:sensitivity_comparison}
\end{table}

The two example sequences (CYL-6 and HoRD-qutrit-8) are likely not optimal sensing sequences, but the physical picture we discussed here provides a convenient method to incorporate quantum sensing into the sequence design procedure.
We can follow the same procedure as described in the preceding sections, but add in maximizing metrological sensitivity as an additional design criteria in choosing the ordering of toggling frames.
We leave the detailed design of such sequences to future work.

\section{Conclusions}
\label{sec:conclusions}

In this work, we introduced a graph-based framework for the design of robust disorder and interaction decoupling sequences in qudit systems, and used this to experimentally demonstrate the first full decoupling of qudit interactions.
In particular, our experiments demonstrate that our robust qutrit disorder and interaction decoupling sequence ``DROID-C3PO" results in a ten-fold improvement in coherence time over existing sequences, highlighting the power of our design framework.
This framework only requires tracking the transformation of the $S^z$ operator under pulses (i.e. ``frames"), significantly reducing the sequence search space compared to prior approaches.
Furthermore, by keeping track of all experimentally-implementable connections between frames, we reduced the sequence construction into a simple graph traversal problem, avoiding the complicated, unstructured algebraic simplifications in prior approaches.
Finally, we showed how pulses that transform $S^z$ along geodesics lead to the natural and elegant incorporation of robustness considerations into our framework.

Our work also opens up new opportunities for future studies.
For quantum many-body physics, higher spins enable new classes of Hamiltonians and phenomena, including quantum many-body scars~\cite{schecter2019weak}, new spin-exchange channels~\cite{davis2019photon,stamper-kurn2013spinor}, lattice gauge theories~\cite{gonzalez-cuadra2022hardware,van2022dynamical}, and $SU(N)$-magnetism~\cite{gorshkov2010two,zhang2014spectroscopic}.
With larger Hilbert space dimension, it also becomes possible to detect the Berry phase on a subsystem by using the additional levels as a phase reference.
This may enable the study of interesting topological phenomenona in Floquet engineered systems~\cite{yao2012topological}.
In quantum metrology, the larger spin translates to a larger dipole moment for enhanced sensing~\cite{fang2013high,mamin2014multipulse,bauch2018ultralong}, and in our experimental platform of interacting NV ensembles, using the full spin-1 degree of freedom allows time-reversal operations that are not readily accessible with two levels~\cite{choi2017dynamical}, crucial for entanglement-enhanced metrology~\cite{davis2016approaching,hosten2016quantum} and measurements of out-of-time-ordered-correlators (OTOCs)~\cite{garttner2017measuring}.
In quantum computation, where the use of qudits may have some advantages over qubits in gate complexity~\cite{ralph2007efficient}, our decoupling sequence can be applied to preserve quantum information for longer timescales~\cite{cohen2021quantum} and allow for more quantum operations within the coherence time.
Finally, as a generic framework, our method can be used to design practical decoupling or Hamiltonian engineering sequences for a wide range of experimental platforms, even if they have different dominant decoherence channels or spin greater than 1.
These results may have  wide-ranging implications for a number of different experimental systems beyond NV centers, including quadrupolar NMR~\cite{vega1976fourier,brauniger2013solid,Chandrakumar1996}, cold molecules~\cite{bohn2017cold,lepoutre2019out}, and nuclear spins or hyperfine states in trapped atoms~\cite{patscheider2020controlling,gorshkov2010two,zhang2014spectroscopic,gabardos2020relaxation,davis2020protecting}.
\newline

We thank J.~Choi, S.~Choi, C.~Hart, W.~W.~Ho, N.~Maskara, J.~T.~Oon, H.~Pichler, C.~Ramanathan, Q.-Z.~Zhu for helpful discussions. This work was supported in part by CUA, NSSEFF, ARO MURI, DARPA DRINQS, Moore Foundation GBMF-4306, NSF PHY-1506284.

\bibliography{main}

\end{document}

\title{Supplementary Materials for \\``Robust Hamiltonian Engineering for Interacting Qudit Systems"}

\affiliation{Department of Physics, Harvard University, Cambridge, Massachusetts 02138, USA}
\affiliation{School of Engineering and Applied Sciences, Harvard University, Cambridge, Massachusetts 02138, USA}

\author{Hengyun Zhou$^1$}
\thanks{These authors contributed equally to this work}
\author{Haoyang Gao$^1$}
\thanks{These authors contributed equally to this work}
\author{Nathaniel T. Leitao$^1$}
\author{Oksana Makarova$^{1,2}$}
\author{Iris Cong$^1$}
\author{Alexander M. Douglas$^1$}
\author{Leigh S. Martin$^1$}
\author{Mikhail D. Lukin$^1$}
\email{Corresponding email: lukin@physics.harvard.edu}
\maketitle

\tableofcontents
\section{Definitions and Proofs}
\label{sec:derivation}

\subsection{Convention for Gell-Mann Matrices}
\label{sec:gellmann}
For qutrit Hamiltonians, we adopt the following convention for the Gell-Mann basis, where the basis elements are defined as
\begin{align}
&\lambda_1=\left(
\begin{array}{ccc}
 0 & 1 & 0 \\
 1 & 0 & 0 \\
 0 & 0 & 0 \\
\end{array}
\right),\quad \lambda_2=\left(
\begin{array}{ccc}
 0 & -i & 0 \\
 i & 0 & 0 \\
 0 & 0 & 0 \\
\end{array}
\right),\quad \lambda_3=\left(
\begin{array}{ccc}
 1 & 0 & 0 \\
 0 & -1 & 0 \\
 0 & 0 & 0 \\
\end{array}
\right),\quad \lambda_4=\left(
\begin{array}{ccc}
 0 & 0 & 1 \\
 0 & 0 & 0 \\
 1 & 0 & 0 \\
\end{array}
\right),\nonumber\\
&\lambda_5=\left(
\begin{array}{ccc}
 0 & 0 & -i \\
 0 & 0 & 0 \\
 i & 0 & 0 \\
\end{array}
\right),\quad\lambda_6=\left(
\begin{array}{ccc}
 0 & 0 & 0 \\
 0 & 0 & 1 \\
 0 & 1 & 0 \\
\end{array}
\right),\quad \lambda_7=\left(
\begin{array}{ccc}
 0 & 0 & 0 \\
 0 & 0 & -i \\
 0 & i & 0 \\
\end{array}
\right),\quad \lambda_8=\frac{1}{\sqrt{3}}\left(
\begin{array}{ccc}
 1 & 0 & 0 \\
 0 & 1 & 0 \\
 0 & 0 & -2 \\
\end{array}
\right),\nonumber\\&\lambda_0=\sqrt{\frac{2}{3}}\left(
\begin{array}{ccc}
 1 & 0 & 0 \\
 0 & 1 & 0 \\
 0 & 0 & 1 \\
\end{array}
\right),
\end{align}

\subsection{Constraints on Frame Matrices}
Generically, we can decompose the transformed $\tilde{S}^z$ operator in the generalized Gell-Mann basis $\{\lambda_\mu\}$ as $\tilde{S}^z_k=U_{k-1}^\dagger S^z U_{k-1}=\sum_\mu F_{\mu,k}\lambda_\mu$.

The frame matrices $F_{\mu,k}$ are required to satisfy certain constraints due to the frame transformation $U^\dagger S^z U$ being unitary.
For example, for qubit systems, we require
\begin{align}
    F_{x,k}^2+F_{y,k}^2+F_{z,k}^2=1
\end{align}
for all $k$.

For qudit systems, the constraint is that a unitary conjugation leaves the eigenvalues unchanged, such that $U_{k-1}^\dagger S^z U_{k-1}=\sum_\mu F_{\mu,k}\lambda_\mu$ should have the same set of eigenvalues as $S^z$.
As an example, for qutrit systems, this imposes the requirements
\begin{align}
&F_1^2+F_2^2+F_3^2+F_4^2+F_5^2+F_6^2+F_7^2+F_8^2=1,\\
&\frac{2 F_8^3}{3 \sqrt{3}}-\frac{2 F_1^2 F_8}{\sqrt{3}}-\frac{2 F_2^2 F_8}{\sqrt{3}}-\frac{2 F_3^2 F_8}{\sqrt{3}}+\frac{F_4^2 F_8}{\sqrt{3}}+\frac{F_5^2 F_8}{\sqrt{3}}\nonumber\\&\quad+\frac{F_6^2
   F_8}{\sqrt{3}}+\frac{F_7^2 F_8}{\sqrt{3}}-F_3 F_4^2-F_3 F_5^2+F_3 F_6^2+F_3 F_7^2\nonumber\\&\quad-2 F_1 F_4 F_6-2 F_2 F_5 F_6+2 F_2 F_4 F_7-2 F_1 F_5 F_7=0,
\end{align}
where we have dropped the $k$ index for simplicity.

\subsection{Proof of Theorem~III.1}
\label{sec:seculartheorem}
We can rewrite the condition $U_1^\dagger S^z U_1=U_2^\dagger S^z U_2$ as $[U_1U_2^\dagger,S^z]=0$, which implies that $S^z$ and $U_1U_2^\dagger$ are simultaneously diagonalizable.
Since $S^z$ has non-degenerate eigenvalues, this implies that the matrix $U_1U_2^\dagger$ must also be diagonal.
Moreover, the unitarity of $U_1$ and $U_2$ imply that $U_1U_2^\dagger$ is also unitary.
Combined, these imply that $U_1U_2^\dagger$ is a composition of rotations around the $\hat{z}$-axis of individual transitions.

On the other hand, by definition, $H$ is a qudit Hamiltonian satisfying the secular approximation on each transition, i.e. for any spin rotation $U_z$ formed by the composition of rotations around the $\hat{z}$-axis of individual transitions, we have
\begin{align}
(U_z^\dagger)^{\otimes n}HU_z^{\otimes n}=H.
\end{align}
Since $U_1U_2^\dagger$ belongs to the type of rotation $U_z$, we have
\begin{align}
(U_2U_1^\dagger)^{\otimes n}H(U_1U_2^\dagger)^{\otimes n}=H,
\end{align}
which implies
\begin{align}
(U_1^\dagger)^{\otimes n} H U_1^{\otimes n}=(U_2^\dagger)^{\otimes n} H U_2^{\otimes n}.
\end{align}

\subsection{Proof of Polynomial Representation of Hamiltonian}
\label{sec:poly}
In the case of spin-$\frac{1}{2}$ particles with two-body interactions, we have shown that the transformed Hamiltonian is a second-order polynomial in the coefficients of the Pauli decomposition of $U^\dagger S^zU$. In the more general case, we can generalize this as follows:
\begin{theorem}
For a spin-$\left(\frac{d-1}{2}\right)$ (where $d$ is the dimension of the local Hilbert space) secular Hamiltonian consisting of at most $k$-body terms, if the $S^z$ operator is transformed as $U^\dagger S^zU=\sum_\mu a_\mu\lambda_\mu$, then the transformed Hamiltonian can be written as an order-$k(d-1)$ polynomial in $a_\mu$.
\label{thm:polynomial}
\end{theorem}
\begin{proof}
First, we note that if the transformation of the $S^z$ operator is specified, then so are the transformations of all powers of $S^z$, since $U^\dagger (S^z)^m U=(U^\dagger S^z U)^m$. In addition, linear combinations of the zeroth to $(d-1)$-th powers of the spin-$\left(\frac{d-1}{2}\right)$ operator $S^z$ generate all diagonal matrices of dimension $d$. To generate off-diagonal matrices, we can make use of the spin-$\left(\frac{d-1}{2}\right)$ $S^x$ operator and its powers: the product of diagonal matrices and $S^x$ generates all matrices containing only elements on the $1$-diagonal (i.e. the diagonal that is 1 element offset from the main diagonal), the product of diagonal matrices and $(S^x)^2$ generates all matrices containing only elements on the $2$-diagonal and so on.

Therefore, if we write the transformed spin operators as $\tilde{S}^z=U^\dagger S^zU=\sum_\mu a_\mu\lambda_\mu$ and $\tilde{S}^x=U^\dagger S^xU=\sum_\mu b_\mu\lambda_\mu$, then any other spin operator can be expressed as $\sum_\alpha f_\alpha(a_\mu,b_\mu)\lambda_\alpha$, with $f_\alpha(a_\mu,b_\mu)$ being a multivariate polynomial in $a_\mu$, $b_\mu$ that is at most order $(d-1)$ in each $a_\mu$. The Hamiltonian, consisting of at most $k$-body terms, can in turn be written as a polynomial
\begin{align}
\tilde{H}=\sum_{\alpha_1\cdots\alpha_k}g_{\alpha_1\cdots\alpha_k}(a_\mu,b_\mu)\lambda_{\alpha_1}\otimes\cdots\otimes\lambda_{\alpha_k},
\end{align}
where $g_{\alpha_1\cdots\alpha_k}(a_\mu,b_\mu)$ is a multivariate polynomial in $a_\mu$, $b_\mu$ that is at most order $k(d-1)$ in each $a_\mu$.

At this point, our expression of the Hamiltonian still depends on both $a_\mu$ and $b_\mu$. However, based on Theorem~III.1, we know that this can be simplified into a form that only contains $a_\mu$. The constraints that can be used to perform this simplification are the preservation of eigenvalues (and thus the characteristic polynomial) under unitary transformations, e.g. $|\lambda I-\sum_\mu a_\mu\lambda_\mu|=|\lambda I-S^z|$. This will give rise to polynomial constraint equations within $a_\mu$ values as well as between $a_\mu$ and $b_\mu$. Crucially, the imposition of these to simplify $g_{\alpha_1\cdots\alpha_k}(a_\mu,b_\mu)$ can only lead to a reduction of the order of the polynomial in $a_\mu$. Therefore, we see that the final expression for the Hamiltonian in terms of $a_\mu$ will be a polynomial of order at most $k(d-1)$.
\end{proof}
Explicit expressions for the polynomial can be elegantly obtained using representation theory of Lie groups, as described in more detail in an accompanying paper, Ref.~\cite{leitao2022qudit}.

\subsection{Implementation of Rotations Regardless of Pulse History}
\label{sec:implementability}
The graphical representation we developed in Sec.~III.2 of the main text uses edges to denote whether two $\tilde{S}^z$ frames can be connected via a simple, physically implementable pulse in our given pulse set.
This requires us to show that regardless of the history of the pulses applied before, an appropriate unitary can still be found which implements the rotation in the desired way.
We rigorously show that this will always be the case with the following theorem:

\begin{theorem}
Given unitaries $U_1$ and $U_2$, if the frame $S_1=U_1^\dagger S^z U_1$ can be transformed into the frame $S_2=U_2^\dagger S^z U_2$ via some physically-implementable pulse $P$, i.e.
\begin{align}
U_1^\dagger P^\dagger S^z PU_1=U_2^\dagger S^z U_2,
\end{align}
then a transformation between $S_1$ and $S_2$ can always be implemented by a phase-shifted version of $P$, regardless of how operators other than $S^z$ (e.g. $S^x$) are transformed.
\label{thm:pulse_history}
\end{theorem}

\begin{proof}
Any unitary operation that gives rise to the same $S^z$ transformation can be expressed as $U_1'=U_zU_1$, where $U_z$ is a diagonal rotation matrix, since $S_1'=U_1^\dagger U_z^\dagger S^z U_zU_1=S_1$.
Note that this will modify how non-diagonal operators (e.g. $S^x$) are transformed, and subsequently necessitate a different pulse $P'$ to transform the frame from $S_1$ to $S_2$.

In order to transform the original pulse above into something that can also be implemented for this transverse frame configuration, we require
\begin{align}
U_1^\dagger U_z^\dagger (P')^\dagger S^z P'U_zU_1=U_1^\dagger P^\dagger S^z PU_1,
\end{align}
which is satisfied when $P'=U_z P U_z^\dagger$.

Now, let us consider what this conjugation by a diagonal rotation matrix does to a pulsed rotation.
Making use of the fact that
\begin{align}
U_z^\dagger\exp[-iHt]U_z=\sum_{n=0}^\infty U_z^\dagger\frac{(-iHt)^n}{n!}U_z=\exp[-i(U_z^\dagger HU_z)t],
\end{align}
we see that in order to implement the rotation, we simply need to conjugate the rotation axis by the same diagonal rotation matrix $U_z$.
Importantly, the only effect of such a conjugation is to change the phase of the pulse applied on each transition, without changing which transitions have nonzero amplitudes applied.
This ensures that the resulting pulse is still physically implementable.
\end{proof}
As a concrete example, for a spin-1 system, a generic rotation generator matrix is transformed by a diagonal rotation matrix $U_z=\textrm{diag}\{e^{i\theta_1},e^{-i\theta_1-i\theta_2},e^{i\theta_2}\}$ as
\begin{align}
&U_z\begin{pmatrix}
a_{11} & a_{12} & a_{13} \\ a_{21} & a_{22} & a_{23} \\ a_{31} & a_{32} & a_{33}\\
\end{pmatrix}U_z^\dagger\nonumber\\&=
\begin{pmatrix}
a_{11} & a_{12}e^{-i(2\theta_1+\theta_2)} & a_{13}e^{i(\theta_2-\theta_1)} \\ a_{21}e^{i(2\theta_1+\theta_2)} & a_{22} & a_{23}e^{i(\theta_1+2\theta_2)} \\ a_{31}e^{i(\theta_1-\theta_2)} & a_{32}e^{-i(\theta_1+2\theta_2)} & a_{33}\\
\end{pmatrix}.
\end{align}
As one can see, the only effect is to change the phases of the drives applied on each transition, which does not change the experimental implementability of the pulses.

\section{Details of Pulse Sequences and Frame Sets}
\label{sec:framedetails}

\subsection{Details of Qutrit Decoupling Frame Sets}
\label{sec:decouplingframes}
In this section, we utilize the linear programming formulation described in Ref.~\cite{choi2017dynamical} to identify promising candidates frame sets.

Before discussing the details, let us first comment on the existence of some equivalence relations between distinct frame sets, in close relation to the results in Appendix.~\ref{sec:implementability}.
Consider the frames formed by conjugating by pulses $P_1$ and $P_2$:
\begin{align}
S_1=P_1^\dagger S^z P_1,\quad S_2=P_1^\dagger P_2^\dagger S^z P_2P_1,
\end{align}
then a further conjugation by $U$ would give rise to
\begin{align}
S_1'=U^\dagger P_1^\dagger S^z P_1 U,\quad  S_2'=U^\dagger P_1^\dagger P_2^\dagger S^z P_2P_1U.
\end{align}
However, physically speaking, we could regard $U$ as an initial state preparation pulse.
The subsequent decoupling pulses will then be unaffected, and therefore from the perspective of the average Hamiltonian, the two sequences are equivalent.

Given this equivalence, when performing calculations, it may be convenient to conjugate the whole pulse set by the same unitary $P_1^\dagger$, in order to start with the frame $S^z$.
This will also allow us to analyze different pulse sequences on a more equal footing.
Moreover, this helps to prevent potential confusions in analyzing pulse sequences related to the order of conjugations when applying a pulse sequence. For example, it may appear that the Gell-Mann basis $\lambda_1$ can be transformed into $\lambda_4$ by a cyclic echo pulse $P_c$ (see \ref{sec:robustdisorder} for definition), in the sense that $P_c^\dagger\lambda_1 P_c=\lambda_4$; but in fact, if one starts in the frame $S_1=\lambda_1=P_1^\dagger S^z P_1$, then the cyclic echo pulse $P_c$ will not bring one to the frame $\lambda_4$, i.e. $P_1^\dagger P_c^\dagger S^z P_cP_1\neq\lambda_4$, since the new pulse acts from the middle instead of being added at the ends.

After imposing that one starts with the frame $S^z$, there is still an additional degree of freedom to conjugate pulse sequences, namely a conjugation of all frames by a diagonal phase rotation.
This will give rise to a family of frame configurations that are distinct, but can be related to each other.
Moreover, if one member of this family can be implemented with some set of elementary pulses composed of resonant driving on one or both magnetically-allowed transitions, the pulses necessary to implement a different member of this family can be easily obtained by changing the phase of the original pulses.
This could be useful in performing additional symmetrization of a pulse sequence to further improve its performance and cancel higher-order terms.

We now describe some of the promising frame sets that were found.
By allowing $\pi/2$ driving pulses on all 3 transitions, including the magnetically-forbidden transition, we found the following 12 frames with equal time duration achieves full disorder and interaction decoupling:
\begin{align}
    \left(
\begin{array}{ccc}
 0 & 0 & 0 \\
 0 & 0 & \pm 1 \\
 0 & \pm 1 & 0 \\
\end{array}
\right), \left(
\begin{array}{ccc}
 0 & 0 & 0 \\
 0 & 0 & \pm i \\
 0 & \mp i & 0 \\
\end{array}
\right), \left(
\begin{array}{ccc}
 0 & 0 & \pm 1 \\
 0 & 0 & 0 \\
\pm 1 & 0 & 0 \\
\end{array}
\right),\nonumber\\ \left(
\begin{array}{ccc}
 0 & 0 & \pm i \\
 0 & 0 & 0 \\
\mp i & 0 & 0 \\
\end{array}
\right), \left(
\begin{array}{ccc}
 0 & \pm 1 & 0 \\
 \pm 1 & 0 & 0 \\
 0 & 0 & 0 \\
\end{array}
\right), \left(
\begin{array}{ccc}
 0 & \pm i & 0 \\
 \mp i & 0 & 0 \\
 0 & 0 & 0 \\
\end{array}
\right).
\end{align}
This frame set does not contain $S^z$ as an element.
Therefore, based on the preceding discussion, we can perform a global unitary rotation to bring this frame set into the following form (alternatively, this can also be directly found by performing linear programming with a pulse set composed purely of balanced double-driving pulses):
\begin{align}
    \pm S^x&=\pm\frac{1}{\sqrt{2}}\left(
\begin{array}{ccc}
 0 & 1 & 0 \\
 1 & 0 & 1 \\
 0 & 1 & 0 \\
\end{array}
\right), &
\pm S^{\tilde{x}}&=\pm\frac{1}{\sqrt{2}}\left(
\begin{array}{ccc}
 0 & 1 & 0 \\
 1 & 0 & -1 \\
 0 & -1 & 0 \\
\end{array}
\right),\nonumber\\
\pm S^y&=\pm\frac{1}{\sqrt{2}}\left(
\begin{array}{ccc}
 0 & -i & 0 \\
 i & 0 & -i \\
 0 & i & 0 \\
\end{array}
\right), &
\pm S^{\tilde{y}}&=\pm\frac{1}{\sqrt{2}}\left(
\begin{array}{ccc}
 0 & -i & 0 \\
 i & 0 & i \\
 0 & -i & 0 \\
\end{array}
\right),\nonumber\\
\pm S^z&=\pm\left(
\begin{array}{ccc}
 1 & 0 & 0 \\
 0 & 0 & 0 \\
 0 & 0 & -1 \\
\end{array}
\right), &
\pm S^{\tilde{z}}&=\pm\left(
\begin{array}{ccc}
 0 & 0 & -i \\
 0 & 0 & 0 \\
 i & 0 & 0 \\
\end{array}
\right).
\end{align}
This is the basic frame set that we use for the majority of our qutrit decoupling pulse sequences.
Note that by globally changing the phases of all rotations, we can also generate other equivalent classes of frames. Also note that another way to specify these frames is to specify them as commutators and anti-commutators of spin-1 operators:
\begin{align}
    S^{\left(x,y,z\right)} &\propto \left[S^\mu, S^\nu\right],\nonumber\\
    S^{\left(\tilde{x},\tilde{y},\tilde{z}\right)} &\propto \left\{S^\mu, S^\nu\right\},
\end{align}
with $\mu,\nu\in\{x,y,z\}$, which generate the irreducible representations of $SU(2)$ discussed in Ref.~\cite{leitao2022qudit}.

Another example frame set that was identified with these methods is:
\begin{align}
    \left(
\begin{array}{ccc}
 0 & \frac{1}{\sqrt{2}} & 0 \\
 \frac{1}{\sqrt{2}} & 0 & -\frac{1}{\sqrt{2}} \\
 0 & -\frac{1}{\sqrt{2}} & 0 \\
\end{array}
\right), \left(
\begin{array}{ccc}
 0 & \frac{i}{\sqrt{2}} & 0 \\
 -\frac{i}{\sqrt{2}} & 0 & -\frac{i}{\sqrt{2}} \\
 0 & \frac{i}{\sqrt{2}} & 0 \\
\end{array}
\right), \left(
\begin{array}{ccc}
 -\frac{1}{2} & 0 & -\frac{1}{2} \\
 0 & 1 & 0 \\
 -\frac{1}{2} & 0 & -\frac{1}{2} \\
\end{array}
\right),\nonumber\\
\left(
\begin{array}{ccc}
 0 & \frac{i}{\sqrt{2}} & 0 \\
 -\frac{i}{\sqrt{2}} & 0 & \frac{1}{\sqrt{2}} \\
 0 & \frac{1}{\sqrt{2}} & 0 \\
\end{array}
\right), \left(
\begin{array}{ccc}
 0 & \frac{i}{\sqrt{2}} & 0 \\
 -\frac{i}{\sqrt{2}} & 0 & -\frac{1}{\sqrt{2}} \\
 0 & -\frac{1}{\sqrt{2}} & 0 \\
\end{array}
\right), \left(
\begin{array}{ccc}
 -\frac{1}{2} & 0 & -\frac{i}{2} \\
 0 & 1 & 0 \\
 \frac{i}{2} & 0 & -\frac{1}{2} \\
\end{array}
\right),\nonumber\\
\left(
\begin{array}{ccc}
 0 & \frac{i}{\sqrt{2}} & 0 \\
 -\frac{i}{\sqrt{2}} & 0 & \frac{i}{\sqrt{2}} \\
 0 & -\frac{i}{\sqrt{2}} & 0 \\
\end{array}
\right), \left(
\begin{array}{ccc}
 0 & -\frac{i}{\sqrt{2}} & 0 \\
 \frac{i}{\sqrt{2}} & 0 & \frac{i}{\sqrt{2}} \\
 0 & -\frac{i}{\sqrt{2}} & 0 \\
\end{array}
\right),\left(
\begin{array}{ccc}
 0 & 0 & 0 \\
 0 & 1 & 0 \\
 0 & 0 & -1 \\
\end{array}
\right), \nonumber\\ \left(
\begin{array}{ccc}
 0 & -\frac{i}{\sqrt{2}} & 0 \\
 \frac{i}{\sqrt{2}} & 0 & -\frac{1}{\sqrt{2}} \\
 0 & -\frac{1}{\sqrt{2}} & 0 \\
\end{array}
\right), \left(
\begin{array}{ccc}
 0 & -\frac{i}{\sqrt{2}} & 0 \\
 \frac{i}{\sqrt{2}} & 0 & \frac{1}{\sqrt{2}} \\
 0 & \frac{1}{\sqrt{2}} & 0 \\
\end{array}
\right), \left(
\begin{array}{ccc}
 -1 & 0 & 0 \\
 0 & 1 & 0 \\
 0 & 0 & 0 \\
\end{array}
\right),
\end{align}
but we do not use this frame set in practice because its graph connectivity is considerably worse.

\subsection{Details of Qutrit Decoupling Pulse Sequences}
\label{sec:decouplingsequences}
In this section, we will describe in detail the sequences Seq.~2 (Interaction Decoupling), Seq.~B (Non-Robust Decoupling) and Seq.~C (DROID-C3PO) we mentioned in Fig.~8(a).

\begin{figure}
\begin{center}
\includegraphics[width=0.6\columnwidth]{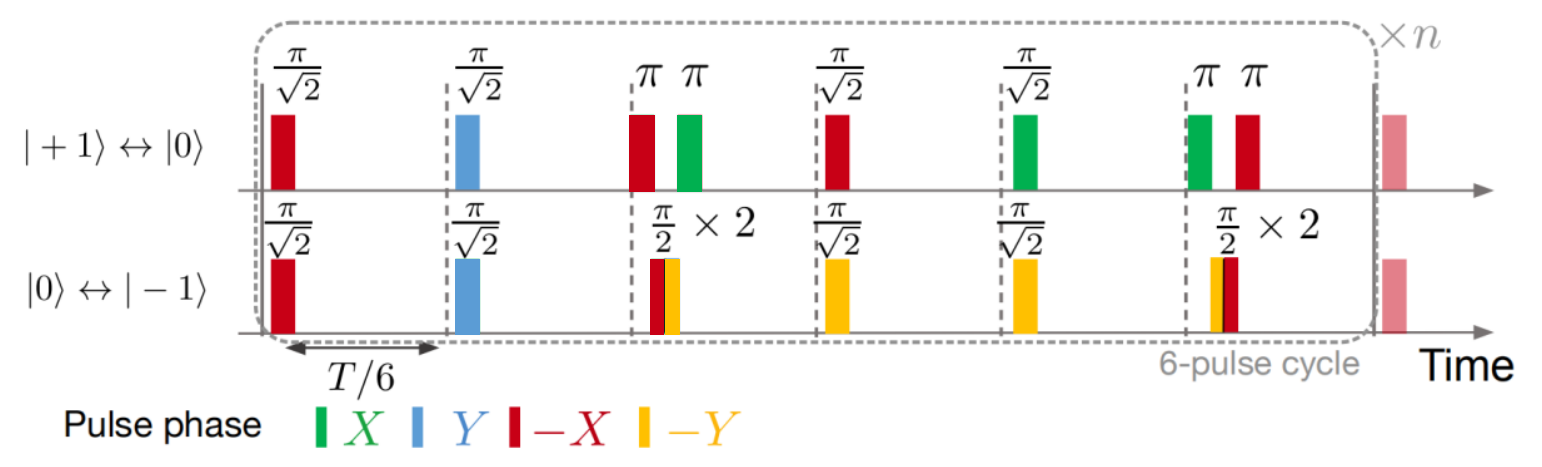}
\caption{{\bf The pulse sequence ``Interaction Decoupling".} This sequence was proposed in Ref.~\cite{choi2017dynamical}.}
\label{fig:Soonwon_Sequence}
\end{center}
\end{figure}

Seq.~2 (Interaction Decoupling) is the interaction decoupling sequence designed in Ref.~\cite{choi2017dynamical}. This sequence only decouples interaction but not disorder, so its performance is not expected to be good in our experimental platform of interacting NV ensembles, because our system is disorder-dominated. This pulse sequence is plotted in Fig.~\ref{fig:Soonwon_Sequence}.

\begin{figure}
\begin{center}
\includegraphics[width=0.6\columnwidth]{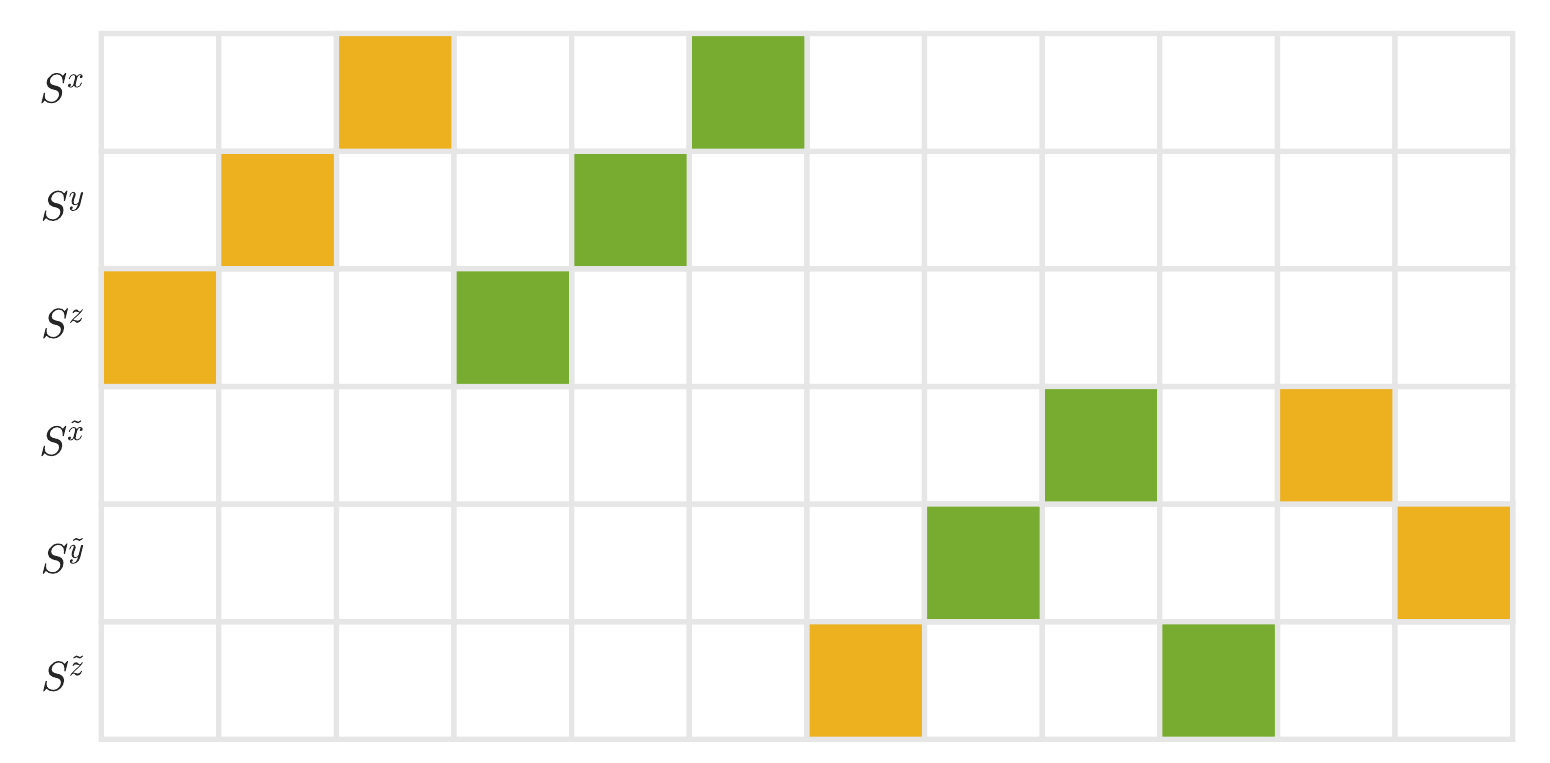}
\caption{{\bf The frame representation of Seq.~B (Non-Robust Decoupling).} This sequence is the sequence plotted in Fig.~3(e-f), which is a non-robust disorder and interaction decoupling sequence.}
\label{fig:Non-robust interaction decoupling}
\end{center}
\end{figure}

Seq.~B (Non-Robust Decoupling) is the sequence plotted in Fig.~3(e-f), which go through the 12 frames in a somewhat arbitrary fashion. When spending equal time in the 12 frames, it is a disorder and interaction decoupling sequence that is not robust to finite pulse duration effects. The frame representation of this sequence is shown in Fig.~\ref{fig:Non-robust interaction decoupling}.

\begin{figure*}
\begin{center}
\includegraphics[width=\columnwidth]{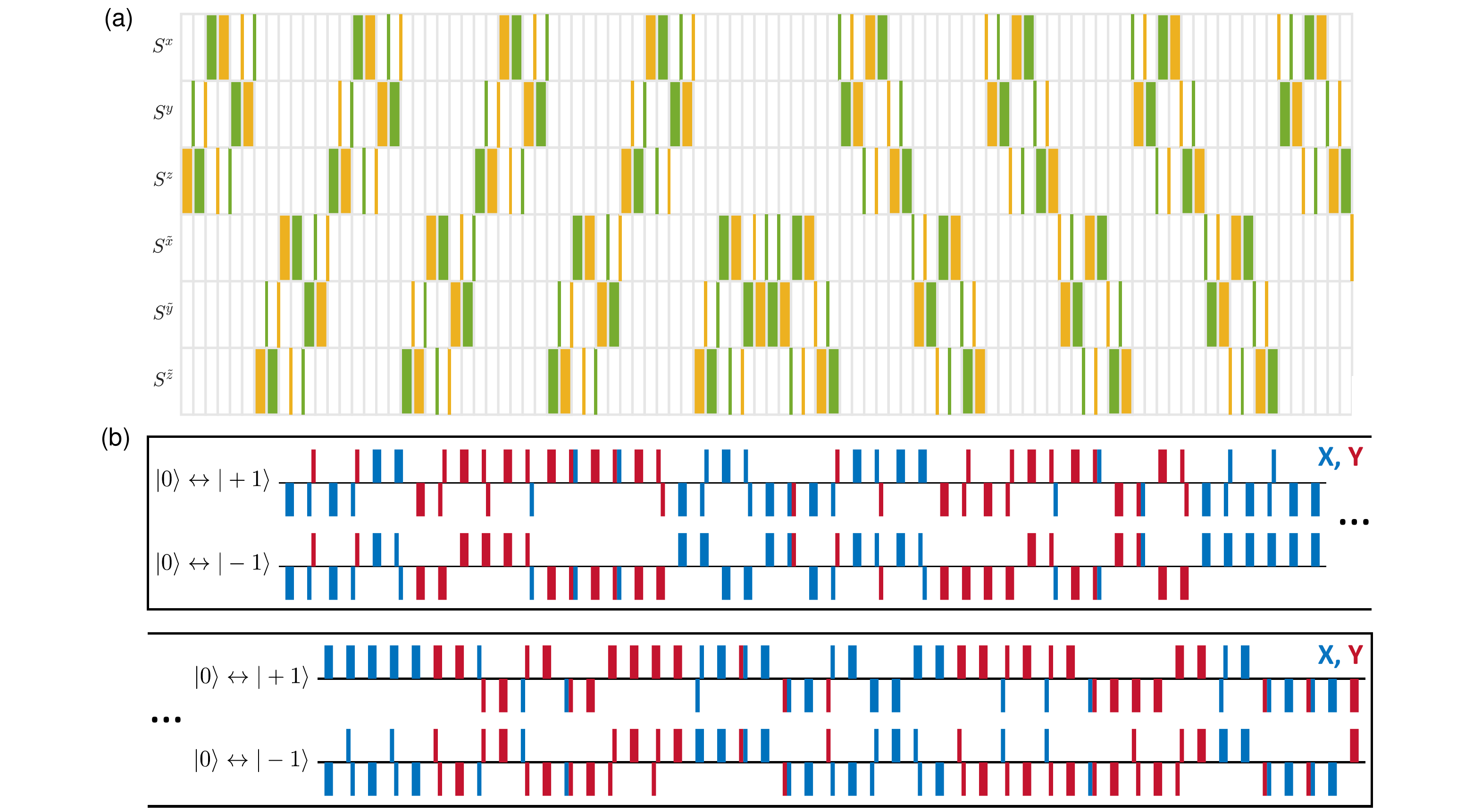}
\caption{{\bf Plot of Seq.~C (DROID-C3PO).} This sequence is our current best disorder and interaction decoupling sequence. It is robust to disorder during pulses, rotation angle errors in both transition, and also cancels some higher order terms in the Magnus expansion. (a) The frame representation of this sequence. (b) The actual pulses constituting this sequence. All pulses in this sequence are balanced double driving. The thin lines represent spin-1 $\pi/2$ pulses (i.e. rotation of the spin-1 generalized Bloch sphere by an angle $\pi/2$, experimentally implemented by simultaneously driving the two transitions with two $\frac{\pi}{\sqrt{2}}$ pulses) and the thick lines represent spin-1 $\pi$ pulses. The color of the pulses represent the pulse axis (X or Y), and the direction of the pulses (up or down) represent the two opposite rotation directions (e.g. $+\pi/2$ pulse and $-\pi/2$ pulse). The proportions of this plot are drawn consistently with actual time durations. The ellipsis in the plot indicates that the two rows are connected. The plot is identical to Fig.~2(f), repeated here for convenience.}
\label{fig:WAHUHA_13}
\end{center}
\end{figure*}

Seq.~C (DROID-C3PO) is our current best sequence whose design is discussed in Sec.~IV of the main text. The sequence in plotted in Fig.~2(f), and its frame representation is shown in Fig.~\ref{fig:WAHUHA_13}(a).

One subtle point about Seq.~C is that it has a net $\pi$ rotation in each Floquet period. Namely, the unitary due to the pulses in each Floquet period is
\begin{align}
    \hat{U} = \left(
\begin{array}{ccc}
 -1 & 0 & 0 \\
 0 & 1 & 0 \\
 0 & 0 & -1 \\
\end{array}
\right).
\end{align}
This net rotation has the potential advantage that the frames in two neighboring Floquet periods are not exactly the same and therefore allows further cancellation between Floquet periods, but it also requires one to be careful because the net rotation changes the readout axis.

\subsection{Robust Disorder Decoupling}
\label{sec:robustdisorder}
In this section, we will describe the disorder decoupling sequences Seq.~1 (Cyclic Echo) and Seq.~A (Robust Cyclic Echo) we mentioned in Fig.~8(a).

\begin{figure}
\begin{center}
\includegraphics[width=0.6\columnwidth]{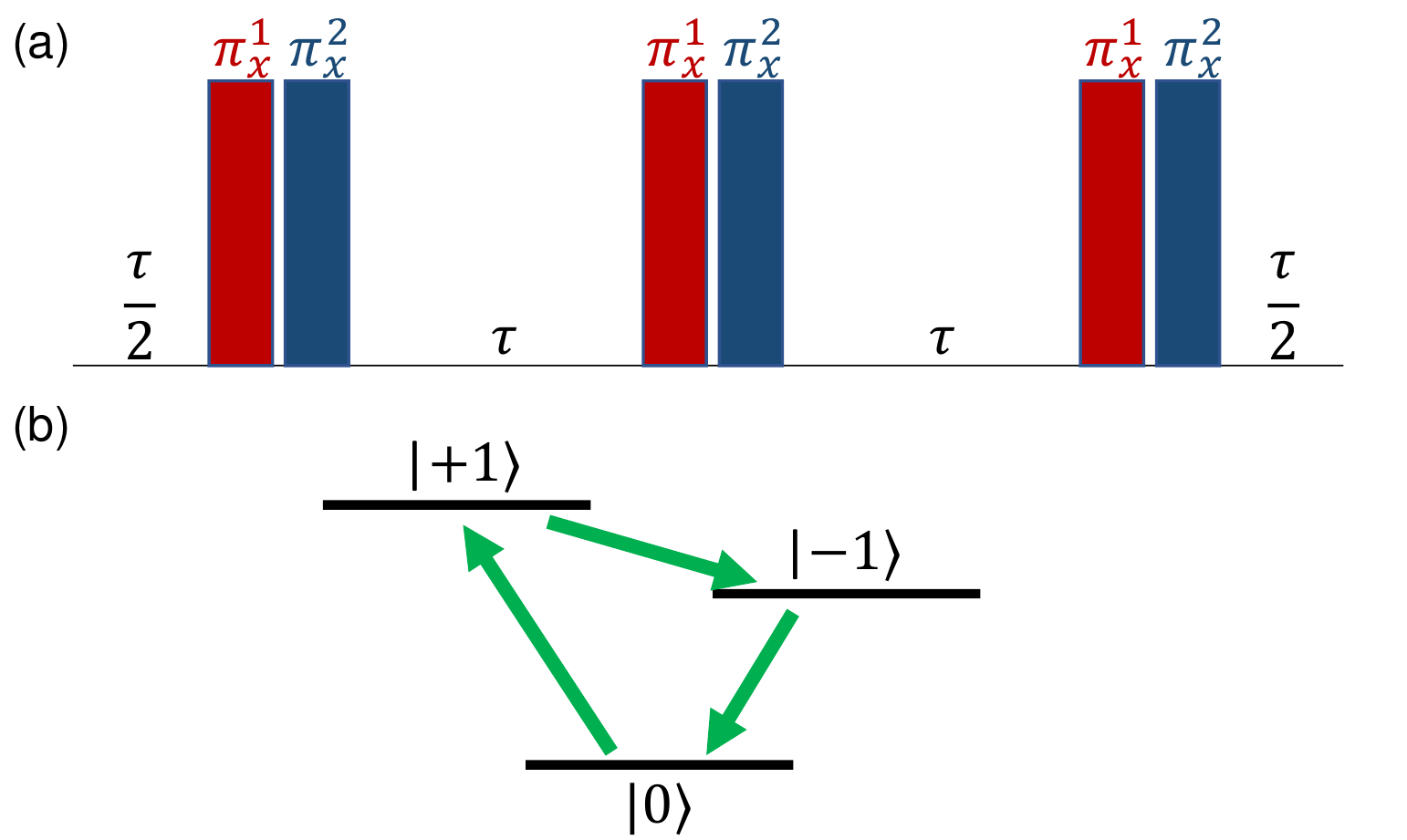}
\caption{{\bf Plot of Cyclic Echo.} (a) The cyclic Echo consists of three pairs of $\pi$ pulses as shown in the plot. The pulses $\pi_x^1$ and $\pi_x^2$ represent a $\pi$ pulse around the $x$ axis for the transition $\ket{0}\leftrightarrow\ket{+1}$ and the transition $\ket{0}\leftrightarrow\ket{-1}$, respectively. (b) Each pair of $\pi$ pulses in (a) causes a cyclic permutation of the three states (as shown by the green arrows) and therefore the disorder is averaged out by this sequence.}
\label{fig:CyclicEcho}
\end{center}
\end{figure}

Seq.~1 (Cyclic Echo) is the simplest sequence that allows one to decouple the on-site disorder. The sequence is plotted in Fig.~\ref{fig:CyclicEcho} and the way it works is to cyclically permutate the three states $\ket{+1}$, $\ket{0}$, and $\ket{-1}$ to average out the disorder.

Seq.~A (Robust Cyclic Echo) is a sequence built on ``Cyclic Echo" that decouples disorder and is robust to finite pulse effects and Rabi inhomogeneity. This sequence consists of two iterations of ``Cyclic Echo" where the phase of pulses in the second iteration is adjusted to $\left(-\pi_x^1, \pi_x^2, \pi_x^1, -\pi_x^2, -\pi_x^1, \pi_x^2\right)$. The way this sequence cancels the disorder during the pulses is by a simple one-to-one cancellation between the two iterations, which is not hard for the readers to verify explicitly. For optimization of performance, we further symmetrized the sequence by adding a free evolution time $\tau$ between the pairs of $\pi$ pulses in Fig.~\ref{fig:CyclicEcho}(a), which changes the cyclic permutation of the three levels shown in Fig.~\ref{fig:CyclicEcho}(b) into a full permutation. This sequence also shares the same net $\pi$ rotation in each Floquet period as discussed in Section.~\ref{sec:decouplingsequences}.

\subsection{Derivation for Disorder Effects During Pulses}
\label{sec:disorderderivation}
In this section, we will discuss how disorder transforms during pulses, which is essential to understand for designing sequences robust to it.

\begin{figure}
\begin{center}
\includegraphics[width=0.6\columnwidth]{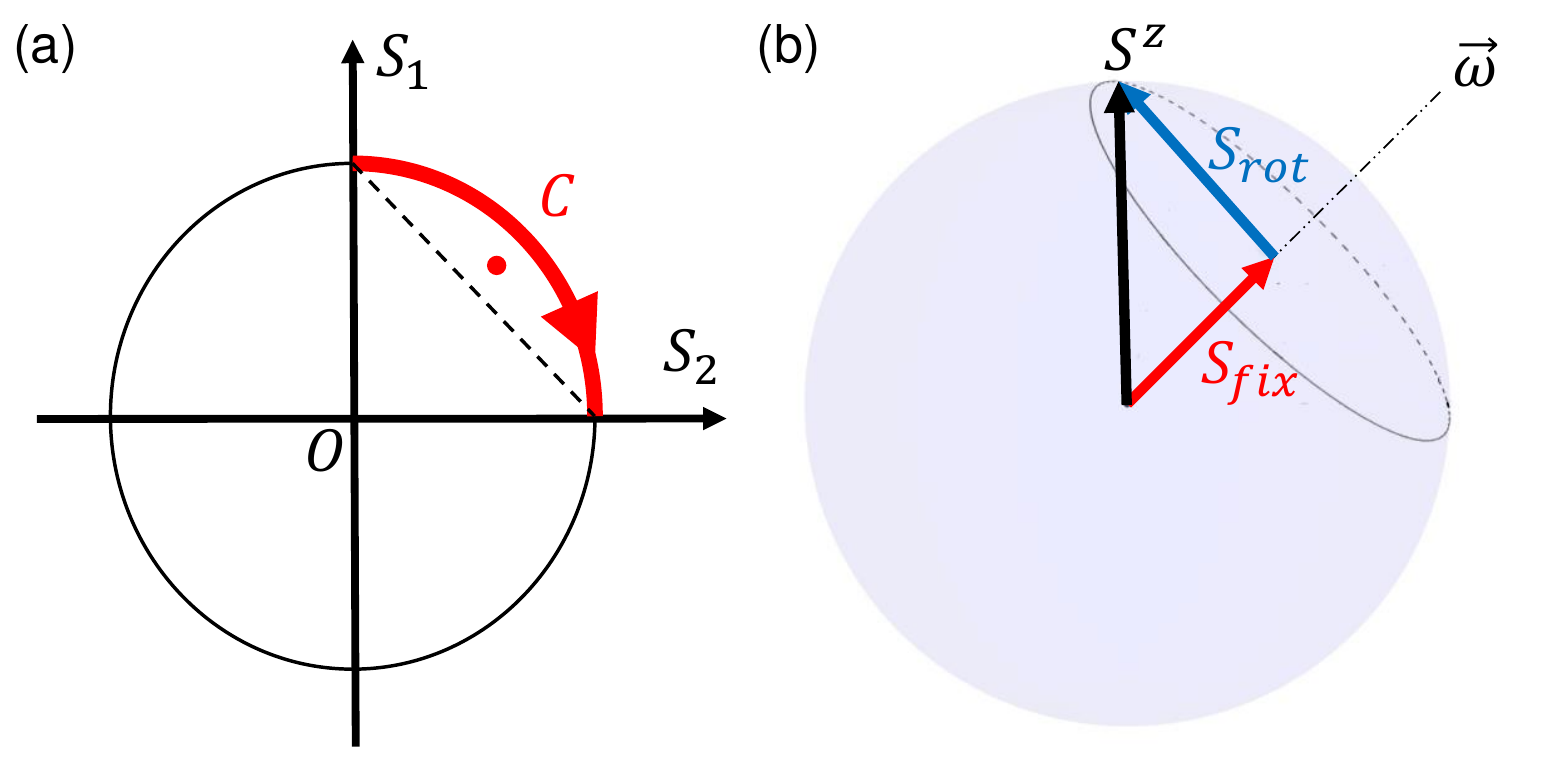}
\caption{{\bf Transformation of $S^z$ during pulses.} (a) When the trajectory of $S^z$ lives on a geodesic, its transformation during the pulse is a simple rotation from its initial position $S_1$ to its final position $S_2$, as shown by the red arc. Then the averaged $S^z$ operator during the pulse is represented by the center of mass of the red arc. From the plot, we can see that the averaged $S^z$ operator can be decomposed as a simple average of $S_1$ and $S_2$, and the extra $\frac{4}{\pi}$ factor in Eq.~(\ref{eq:spin-1/2_averaged_Sz}) comes from the fact that the center of mass of the red curve is slightly further from the origin O compared to the midpoint between $S_1$ and $S_2$. The plot is identical to Fig.~4(a), repeated here for convenience. (b) When the trajectory of $S^z$ lives in a 2 dimensional slice that does not go through the origin (as shown by the black circle in the plot), $\tilde{S}^z\left(t\right)$ can be decomposed into an invariant part $S_{fix}$ during the pulse and a part $S_{rot}$ that rotates on a circle, with the latter's averaged effect follows the same rule as in plot (a).}
\label{fig:Sz_during_pulses}
\end{center}
\end{figure}

Before going into the details for the spin-1 case, let us remind readers of the simple geometric picture in the spin-$\frac{1}{2}$ case.
In the spin-$\frac{1}{2}$ case, as we already discussed in the main text, an on-resonance pulse leads to a $S^z$ operator trajectory that transforms along a geodesic on the Bloch sphere (represented by the red arc in Fig.~4(a) and repeated here in Fig.~\ref{fig:Sz_during_pulses}(a) for convenience). As a result, the averaged effect of disorder during the pulse, as represented by the center of mass of the red arc in Fig.~\ref{fig:Sz_during_pulses}(a), can be decomposed as an average of the frames before and after the pulse:
 \begin{equation}
    \bar{S}=\frac{4}{\pi}\left[\frac{S_1+S_2}{2}\right],
\label{eq:spin-1/2_averaged_Sz}
\end{equation}
where the factor $\frac{4}{\pi}$ comes from the fact that the center of mass is slightly further from the origin than the midpoint between $S_1$ and $S_2$.

Even if the pulse is not on resonance, the story does not change too much because the trajectory of the $S^z$ operator on the Bloch sphere is still a circle (the only difference is that now the circle is not a geodesic on the Bloch sphere). In this case, the evolution of $S^z$ operator can be decomposed into two parts as shown in Fig.~\ref{fig:Sz_during_pulses}(b): the first part is the projection of $S^z$ on the rotation axis, which is invariant during the pulse; and the second part is the remaining part, which rotates on a circle perpendicular to the rotation axis and therefore whose effect during the pulse can be decomposed as a linear combination before and after the pulse, similar to the case in Fig.~\ref{fig:Sz_during_pulses}(a).

Although the geometric picture in the spin-$\frac{1}{2}$ case is very simple (the trajectory of the $S^z$ operator is always a circle), the trajectory of the $S^z$ operator in the spin-1 case can be much more complicated. 
To see this, let us consider the trajectory of the $S^z$ operator transformed by a generic Hamiltonian $H$:
\begin{equation}
    \tilde{S}^z\left(t\right) = e^{iHt}S^z e^{-iHt}.
\label{eq:Sz_during_pulse}
\end{equation}

Generically, the conjugation of the $SU(3)$ operator $e^{-iHt}$ on $S^z$ leads to a rotation on an 8 dimensional sphere which is the spin-1 generalization of the Bloch sphere (the coordinates on the 8 dimensional sphere are the expansion coefficients of $\tilde{S}^z$ in Gell-mann basis, see Ref.~\cite{choi2017dynamical,macfarlane1968description} for more details). In order to analyze the trajectory of $\tilde{S}^z\left(t\right)$ on this 8-dimensional sphere, let us calculate its time derivatives at $t=0$:
\begin{align}
    \tilde{S}^z|_{t=0} &= S^z\nonumber\\
    \frac{d}{dt}\tilde{S}^z|_{t=0} &= \left[iH,S^z\right]\nonumber\\
    \frac{d^2}{dt^2}\tilde{S}^z|_{t=0} &= \left[iH,\left[iH,S^z\right]\right]\label{eq:Sz_derivatives}\\
    ...\nonumber
\end{align}
From the above expression, we know that the trajectory of $\tilde{S}^z\left(t\right)$ lives in a subspace spanned by \{$\left[iH,S^z\right]$, $\left[iH,\left[iH,S^z\right]\right]$, ...\}. Therefore, the number of linearly independent matrices in the set \{$\left[iH,S^z\right]$, $\left[iH,\left[iH,S^z\right]\right]$, ...\} is the dimension of the subspace that the trajectory of $\tilde{S}^z\left(t\right)$ lives in.

By calculating the rank of the above set for randomly chosen $H$, we know that for generic pulses, the dimension of the subspace is 6. Even when we restrict the pulses to be on resonance, the dimension is still 4. Therefore, the trajectory of $\tilde{S}^z\left(t\right)$ is very complicated in the generic case and there is no simple expression for the disorder during pulses. However, we can still get useful results in certain special cases:
\begin{itemize}
    \item Most importantly, as we already discussed in the main text, when the pulse is a balanced double driving pulse (i.e. simultaneously driving the two magnetically allowed transitions with equal amplitude), the trajectory of $S^z$ operator becomes a geodesic, and therefore all nice properties in the spin-$\frac{1}{2}$ case are recovered. This is the key insight that allows us to elegantly cancel the disorder during pulses.
    \item When the pulse is a resonant driving on a single transition, the trajectory of $S^z$ can be decomposed into a fixed part and a rotating part similar to Fig.~\ref{fig:Sz_during_pulses}(b). To see this, consider a pulse applied on the $\ket{0}\leftrightarrow\ket{-1}$ transition. In this case, we can explicitly decompose $S^z$ as:
    \begin{equation}
        S^z=\begin{pmatrix}
           1 & 0 & 0\\
           0 & 0 & 0\\
           0 & 0 & -1\\
           \end{pmatrix}
           =\begin{pmatrix}
           1 & 0 & 0\\
           0 & -\frac{1}{2} & 0\\
           0 & 0 & -\frac{1}{2}\\
           \end{pmatrix}+\begin{pmatrix}
           0 & 0 & 0\\
           0 & \frac{1}{2} & 0\\
           0 & 0 & -\frac{1}{2}\\
           \end{pmatrix},
    \end{equation}
    where the first term is invariant under the rotation, and the second term rotates as a spin-$\frac{1}{2}$ $S^z$ operator during the pulse. If the pulse is a $\frac{\pi}{2}$ pulse, then the average effect of the second term during the pulse is given by Eq.~(\ref{eq:spin-1/2_averaged_Sz}), and the average effect of the first term is simply itself. Because of the extra coefficient $\frac{4}{\pi}$ that only appears for the second term, when the two terms are summed together, their average effect is no longer a simple average before and after the pulse. This is a concrete example that shows the complication in the qudit case compared to qubit case.
    \item Although the trajectory of $\tilde{S}^z$ is very complicated for generic on-resonance pulses, the trajectory of $\left(\tilde{S}^z\right)^2$ always lives in a 2 dimensional space (i.e. looks like the trajectory in Fig.~\ref{fig:Sz_during_pulses}(b)). To see this, notice that when $\ket{0}$ is coupled to $\ket{+1}$ and $\ket{-1}$ by on resonant pulses, it can also be viewed that $\ket{0}$ is coupled to a bright state $\ket{B}$, while leaving a dark state $\ket{D}$ not coupled to anything. For convenience, we can do a basis transformation from $\{\ket{0},\ket{+1},\ket{-1}\}$ to $\{\ket{0},\ket{B},\ket{D}\}$. Since $\left(S^z\right)^2$ is identity in the $\{\ket{+1},\ket{-1}\}$ subspace, it is invariant under this basis transformation. Working in this bright and dark state basis, since $\ket{0}$ is only coupled to $\ket{B}$, the transformation of $\left(S^z\right)^2$ is kept block diagonal, with one block (correspond to $\ket{D}$) invariant and the other block (correspond to $\{\ket{0},\ket{B}\}$) transforming as a two-level-system. Because of this, the trajectory of $\left(\tilde{S}^z\right)^2$ can be decomposed into a fixed part and a rotating part, as shown in Fig.~\ref{fig:Sz_during_pulses}(b)).
    As a specific example, for a spin-1 $\frac{\pi}{2}$ pulse
    \begin{equation}
        U_p=\textrm{exp}[-i\begin{pmatrix}
           0 & \theta_1+i\theta_2 & 0\\
           \theta_1-i\theta_2 & 0 & \theta_3+i\theta_4\\
           0 & \theta_3-i\theta_4 & 0\\
           \end{pmatrix}],
    \label{eq:generic_pulse}
    \end{equation}
    with $\theta_1^2+\theta_2^2+\theta_3^2+\theta_4^2=\theta_{tot}^2=\pi^2/4$, the averaged $\left(S^z\right)^2$ operator during this pulse is
    \begin{equation}
        \bar{S} = \frac{4}{\pi}\left[\frac{\left(S_1^2-S_{fix}\right)+\left(S_2^2-S_{fix}\right)}{2}\right] + S_{fix},
    \label{eq:average_sz_square}
    \end{equation}
    where $S_1$ and $S_2$ are the frames before and after the pulse, and $S_{fix}=\frac{1}{2\pi}\int_0^{2\pi}d\theta U_p^\dagger\left(\theta\right)\left(S^z\right)^2 U_p\left(\theta\right)$ is the invariant part of $\left(S^z\right)^2$ during the pulse.
\end{itemize}

\subsection{Analysis of Rotation Angle Error}
\label{sec:rotationangleerror}
In this section, we will show that the robust qutrit decoupling sequence we designed (see Fig.~\ref{fig:WAHUHA_13}(a) for its frame representation) is not only robust to rotation angle errors common to both transitions, as discussed in the main text, but also robust to rotation angle errors on each individual transition.

To see this, examine Fig.~\ref{fig:WAHUHA_13}(a) and see what we exactly did in the ``Further improvement" level in the hierarchy described in the main text. In the whole sequence ``DROID-C3PO", there are 8 iterations of the basic disorder and interaction decoupling sequence shown in Fig.~5(c). The difference between the first 4 iterations is that signs of free evolution frames and intermediate frames are flipped. For two neighboring frames $A$ and $B$, their signs go over all four possibilities $\left(A,B\right)$, $\left(A,-B\right)$, $\left(-A,B\right)$, and $\left(-A,-B\right)$. The second 4 iterations are obtained by flipping both the signs and the ordering of the frames in the first 4 iterations~\cite{zhou2023robust}. Due to the structure discussed above, for any neighboring frames $\left(A,B\right)$ in the first 4 iterations, there is a pair of frames $\left(-A,-B\right)$ in the first 4 iterations, and therefore there is a pair of frames $\left(B,A\right)$ in the second 4 iterations. Because the rotation from frame $B$ to frame $A$ is exactly the reverse rotation from $A$ to $B$, the rotation angle error on each individual transition is cancelled between frame pairs $\left(A,B\right)$ in the first 4 iterations and $\left(B,A\right)$ in the second 4 iterations. This is how rotation angle errors on each individual transition get cancelled in the sequence.

\subsection{Geometric Intuition of Scar Subspace}
\label{sec:scargeometricpicture}
In this section, we will discuss the geometric structure of the scar subspace $\ket{S_n}$ as defined in Eq.~(12) of the main text. When restricted to the subspace spanned by $\ket{+1}$ and $\ket{-1}$, the operator $\frac{1}{2}\left(S_i^+\right)^2$ becomes the spin-$\frac{1}{2}$ raising operator. If we further rotate the spins in the second group by $\pi$ around the $z$ axis, this raising operator will flip its sign (because the signs of $S^x$ and $S^y$ are flipped) and the operator $J^+$ will become exactly the many-body raising operator. Since the state $\ket{\Omega}$ is the state $\ket{S=\frac{N}{2},m_S=-\frac{N}{2}}$, the states $\ket{S_n}$ will be $\ket{S=\frac{N}{2},m_S=-\frac{N}{2}+n}$ after rotating the spins in the second group by $\pi$ around the $z$ axis. Therefore, the subspace spanned by $\ket{S_n}$ is the maximal spin subspace after rotating the second group by $\pi$ around the $z$ axis.

\subsection{Decoupling with Non-Geodesic Pulses}
\label{sec:nongeodesic}
In this section, we will show another robust qutrit disorder and interaction decoupling sequence whose frame set is different from the 12 frames shown in Fig.~3(d) and whose pulses do not lead to a geodesic trajectory of $\tilde{S}^z$ (see Sec.~III.3 of the main text for the context of geodesics).

The basic idea of this sequence is also a hierarchical design: since disorder is much stronger than interactions in our experimental platform, we want to robustly decouple disorder first, and then decouple interactions on top of that. Therefore, we can use Seq.~A (Robust Cyclic Echo) (see Sec.~\ref{sec:robustdisorder} for descriptions) as the inner layer to robustly decouple the disorder, and on top of that, design sequences to decouple the interaction transformed by ``Robust Cyclic Echo".

\begin{figure}
\begin{center}
\includegraphics[width=0.6\columnwidth]{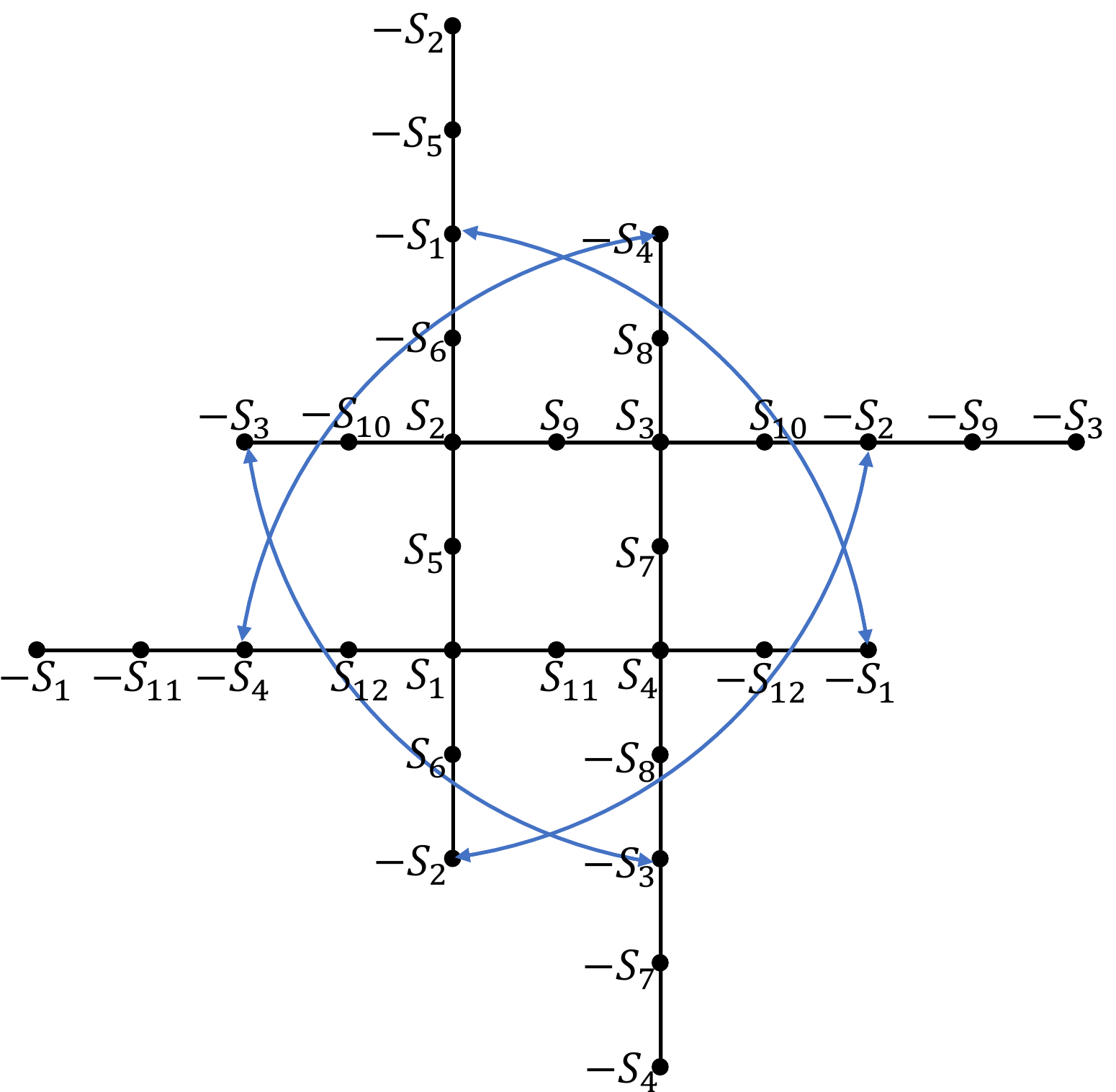}
\caption{{\bf The connectivity graph of the 12 frames in Eq.~(\ref{eq:cyclic_based_frames}).} Each vertex represent one frame and each segment represent a balanced double driving $\frac{\pi}{4}$ pulse. The four straight lines represent four geodesics on which frames can be transformed to each other by repeating the same spin-1 $\frac{\pi}{4}$ pulse. Notice that vertices with the same label represent the same frame. For convenience of readers, 4 pairs of identical frames are connected by blue arcs.}
\label{fig:four_geodesics}
\end{center}
\end{figure}

\begin{figure*}
\begin{center}
\includegraphics[width=\columnwidth]{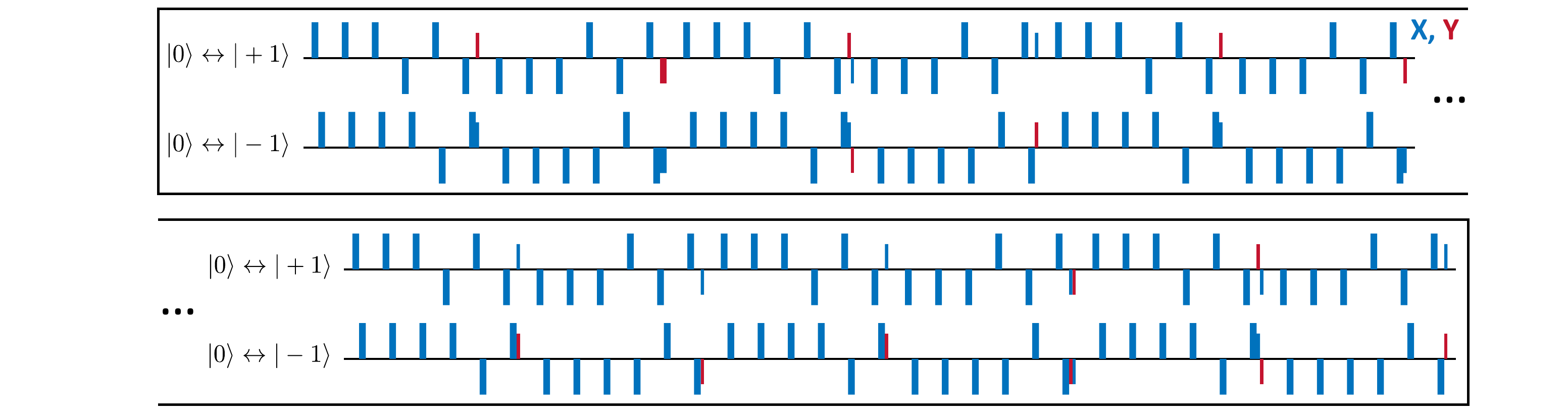}
\caption{{\bf A Non-geodesic decoupling pulse sequence.} There are two different types of pulses in this pulse sequence. One type of pulses are $\pi$ pulses on a single transition, represented by the taller pulses in the plot; these pulses constitute the ``Robust Cyclic Echo", which decouples disorder robustly and locally. The other type of pulses are balanced double driving $\pi/4$ or $\pi/2$ pulses connecting the frames in Fig.~\ref{fig:four_geodesics}, represented by the shorter pulses in the plot; these pulses further decouple interactions on top of disorder decoupling building blocks. The color of the pulses represent the pulse axis (X or Y), and the direction of the pulses (up or down) represent the two opposite rotation directions (e.g. $+\pi/2$ pulse and $-\pi/2$ pulse). The proportion of this plot is drawn consistently with actual time duration. The ellipsis in the plot represent that the two rows are connected.}
\label{fig:CyclicEcho_Based}
\end{center}
\end{figure*}

Because the sequence ``Robust Cyclic Echo" cyclically permutes the three energy levels, the form of the interaction is symmetrized under the transformation of this sequence. More concretely, the original interaction Hamiltonian, which only contains flip-flop terms between $\ket{0}\leftrightarrow\ket{+1}$ and between $\ket{0}\leftrightarrow\ket{-1}$, is transformed to
\begin{align}
    H_{int}^\prime&=\frac{1}{2}S^z\otimes S^z + \frac{1}{2}S_\perp^z\otimes S_\perp^z\nonumber\\ 
    &-\frac{1}{3}H^{XY,+0} - \frac{1}{3}H^{XY,0-} - \frac{1}{3}H^{XY,+-},
\label{eq:symmetrized_interaction}
\end{align}
where $H^{XY,ij}=\frac{1}{2}\left(X^{ij}\otimes X^{ij} + Y^{ij}\otimes Y^{ij}\right)$ is the flip-flop term between state $\ket{i}$ and state $\ket{j}$, and $S_\perp^z$ is defined as
\begin{equation}
    S_\perp^z\equiv\frac{1}{\sqrt{3}}\begin{pmatrix}
           1 & 0 & 0\\
           0 & -2 & 0\\
           0 & 0 & 1\\
           \end{pmatrix}.
\end{equation}
Using the framework we proposed in this paper, we found the following 12 frames with equal time duration decouple the symmetrized interaction Eq.~(\ref{eq:symmetrized_interaction}):
\begin{align}
    S_1 &=\left(
\begin{array}{ccc}
 0 & 1 & 0 \\
 1 & 0 & 0 \\
 0 & 0 & 0 \\
\end{array}
\right), & S_2 &=\left(
\begin{array}{ccc}
 0 & 0 & 0 \\
 0 & 0 & 1 \\
 0 & 1 & 0 \\
\end{array}
\right), & S_3 &=\left(
\begin{array}{ccc}
 0 & -i & 0 \\
 i & 0 & 0 \\
 0 & 0 & 0 \\
\end{array}
\right), & S_4 &=\left(
\begin{array}{ccc}
 0 & 0 & 0 \\
 0 & 0 & -i \\
 0 & i & 0 \\
\end{array}
\right),\nonumber\\ S_5 &=\frac{1}{\sqrt{2}}\left(
\begin{array}{ccc}
 0 & 1 & 0 \\
 1 & 0 & 1 \\
 0 & 1 & 0 \\
\end{array}
\right), & S_6 &=\frac{1}{\sqrt{2}}\left(
\begin{array}{ccc}
 0 & 1 & 0 \\
 1 & 0 & -1 \\
 0 & -1 & 0 \\
\end{array}
\right), & S_7 &=\frac{1}{\sqrt{2}}\left(
\begin{array}{ccc}
 0 & -i & 0 \\
 i & 0 & -i \\
 0 & i & 0 \\
\end{array}
\right), & S_8 &=\frac{1}{\sqrt{2}}\left(
\begin{array}{ccc}
 0 & -i & 0 \\
 i & 0 & i \\
 0 & -i & 0 \\
\end{array}
\right),\nonumber\\ S_9 &=\frac{1}{\sqrt{2}}\left(
\begin{array}{ccc}
 0 & -i & 0 \\
 i & 0 & 1 \\
 0 & 1 & 0 \\
\end{array}
\right), & S_{10} &=\frac{1}{\sqrt{2}}\left(
\begin{array}{ccc}
 0 & -i & 0 \\
 i & 0 & -1 \\
 0 & -1 & 0 \\
\end{array}
\right), & S_{11} &=\frac{1}{\sqrt{2}}\left(
\begin{array}{ccc}
 0 & 1 & 0 \\
 1 & 0 & -i \\
 0 & i & 0 \\
\end{array}
\right), & S_{12} &=\frac{1}{\sqrt{2}}\left(
\begin{array}{ccc}
 0 & 1 & 0 \\
 1 & 0 & i \\
 0 & -i & 0 \\
\end{array}
\right).
\label{eq:cyclic_based_frames}
\end{align}
The connectivity of these 12 frames by experimental pulses is shown in Fig.~\ref{fig:four_geodesics}. As the figure shows, these 12 frames can be connected by balanced double driving (i.e. equal driving amplitude on the two allowed transitions) $\frac{\pi}{4}$ pulses, which is easily implementable in experiments. The whole disorder and interaction decoupling sequence is made of 12 iterations of ``Robust Cyclic Echo", with the double driving $\frac{\pi}{4}$ pulses connecting the above 12 frames inserted between neighboring iterations. A concrete example of such a sequence is given in Fig.~\ref{fig:CyclicEcho_Based}. The sequence in Fig.~\ref{fig:CyclicEcho_Based} is nearly robust to disorder during pulses because the disorder during the pulses in each iteration of the ``Robust Cyclic Echo" is cancelled. Although the disorder during the $\frac{\pi}{4}$ pulses connecting the above 12 frames are not cancelled, it does not matter too much because these pulses take only a very small fraction of time in the whole sequence. Finally, we note that there are at least three ways to further improve the performance of the sequence in Fig.~\ref{fig:CyclicEcho_Based}:
\begin{itemize}
    \item We can further symmetrize the ``Robust Cyclic Echo" by adding a free evolution time $\tau$ between the pair of $\pi$ pulses consist the cyclic echo, just as in Seq.~A.
    \item The sequence can in fact be made fully robust to disorder during pulses: we can compensate for disorder during the balanced double driving $\frac{\pi}{4}$ pulses by slightly adjusting the free evolution times inside neighboring iterations of ``Robust Cyclic Echo".
    \item Nearly all pulses in the sequence in Fig.~\ref{fig:CyclicEcho_Based} are along the $X$ axis, so there is potential space for further improvements by utilizing pulses along $Y$ axis.
\end{itemize}

\bibliography{si}